\documentclass[fleqn,usenatbib,useAMS]{mnras}

\usepackage{graphicx}	
\usepackage{amsmath}	
\usepackage{multicol}        
\usepackage{bm}		
\usepackage{pdflscape}	
\usepackage{ulem}

\newcommand{\kms}{\,km\,s$^{-1}$} 

\usepackage[T1]{fontenc}
\usepackage{ae,aecompl}

\usepackage{newtxtext,newtxmath}

\title[Disappearing Double Peaks in UNAM-KIAS 613]{Spectral Evolution and Transient Broad-Line Features in the Isolated AGN UNAM-KIAS 613}

\author[Cortes-Suárez et al.]{
Edgar Cortes-Suárez$^{1,2,3}$\thanks{E-mail: ecortes@inaoep.mx},
Paola Marziani$^{4,5}$, Héctor Manuel Hernández-Toledo$^{1}$\thanks{E-mail: hector@astro.unam.mx}, Miguel Angel Aragón-Calvo$^{6}$ \and and Castalia Alenka Negrete$^{1}$.
\\
$^{1}$Universidad Nacional Autónoma de México. Instituto de Astronomía. A.P. 70-264, 04510. Ciudad de México, México.\\
$^{2}$Instituto de Astrofísica, Óptica y Electrónica, Luis Enrique Erro 1, Tonantzintla, 72840. Puebla, México\\
$^{3}$Consejo Nacional de Humanidades, Ciencias y Tecnologías, Av. Insurgentes Sur 1582, 03940. Ciudad de México, México\\
$^{4}$INAF, Osservatorio astronomico di Padova, 35122. Padova, Italy\\
$^{5}$Instituto de Astrofísica de Andalucía (IAA-CSIC), 18008. Granada, Spain\\
$^{6}$Universidad Nacional Autónoma de México. Instituto de Astronomía. A.P. 877, 22800. Ensenada, B.C. , México\\
}

\date{Last updated 2024 April 23; in original form 2013 September 5}

\pubyear{{\the\year{}}}

\begin{document}
\label{firstpage}
\pagerange{\pageref{firstpage}--\pageref{lastpage}}
\maketitle

\begin{abstract}
We present multi-epoch optical spectroscopy of the isolated elliptical galaxy UNAM‑KIAS 613, hosting a low-luminosity Type 1 AGN. Analysis of archival Sloan Digital Sky Survey (SDSS) data from 2006 reveals a distinctive double-peaked broad H$\alpha$ profile, tentatively modeled by a relativistic accretion disk. Follow-up observations in 2018  and 2023 show the disappearance of the red and blue wings, leaving only a single-peaked, central broad component. No significant continuum variability is detected in ASAS‑SN and Catalina light curves over 2012–2025, and multi-wavelength data (radio, mid‑IR, X‑ray) confirm a sub‑Eddington, radio‑quiet AGN (Eddington ratio $\approx$ 0.03-0.04, black hole mass $\approx 10^{7.2}$ M$_\odot$). We propose that the  double-peak structure is in reality transient, and arose from a one‑time bipolar outflow event rather than a stable disk or from a Tidal Disruption Event. The mid‑IR SED and radio luminosity place UK 613 on the boundary between AGN and star formation dominance, suggesting residual star formation, while we have found that the isolated environment seems to be prone to the rejuvenation of ellipticals by recent ($\lesssim$ 1 Gyr) cold gas. We also examined its location within the cosmic web  with the aim of identifying possible distinctive effects imprinted on its spectroscopic properties. Ultimately, our results are consistent that UNAM-KIAS 613 might have undergone a ``turn-off'' of the accretion disk emitting region or a transition between a  radiatively-inefficient and radiatively-efficient accretion mode, and highlight the complex interplay of disk, outflow, host processes and environment in low‑accretion, low-black hole mass AGNs, an AGN population still largely unexplored to-date.

\end{abstract}

\begin{keywords}
galaxies: active -- galaxies: elliptical and lenticular, cD -- galaxies: individual: UNAM-KIAS 613 -- line: profiles -- accretion, accretion disks 
\end{keywords}



\begingroup
\let\clearpage\relax
\endgroup
\newpage

\section{Introduction}
\label{S1}

The advent of spectroscopic surveys like SDSS \citep{York2000}, DESI \citep{Dey2019}, Califa \citep{Sebastian2012}, Sami \citep{Croom2012}, Muse \citep{Urrutia2019}, and ongoing SDSS projects  \citep{Bundy2015,Blanton2017,Kollmeier2017} has significantly expanded the census of early-type galaxies exhibiting broad, double-peaked Balmer emission lines in their nuclear spectra. Among members of this class, the elliptical galaxy Arp 102B is widely recognized as a prototype, with its nuclear emission interpreted as originating in the outskirts of an accretion disk. \citet{Halpern1989} presented compelling models of this type, achieving significant success in reproducing the observations, as well as inferring disk parameters such as the inclination angle and inner and outer radii. An identified issue was that the energy released by the disk at those radii was barely sufficient to explain the observed line fluxes. Therefore, additional heating from a photoionizing source, such as radiation from a hot inner disk or a nonthermal nucleus, was proposed. Among other hypotheses, they proposed the existence of ion-supported tori \citep{Chen1989}, which could simultaneously explain the far-infrared emission peaks observed at 25 mum and the structure of other broad-line radio galaxies such as 3C 390.3.

Many monitoring campaigns have been conducted to understand the changing nature of double-peaked Balmer lines. Long-term studies \citep[e.g.,][]{Shapovalova2013,Popovic2014} covering extended periods revealed small but measurable changes in the H$\alpha$ and H$\beta$ profiles. The red-to-blue flux ratio estimates showed periodic oscillations, which \citet{Newman1997} and \citet{Gezari2007} interpreted as due to transient orbiting hot spots in the accretion disk. However, some inconsistencies with a simple symmetric disk model were observed. For example, \citet{Miller1990} and \citet{Sulentic1990} showed that the red peak sometimes exceeded the blue peak, contradicting expectations from a perfectly axisymmetric relativistic disks (most likely an oversimplification).  

In an effort to find statistically significant numbers of Double-Peaked emitters \citet{Eracleous1994,Eracleous2003}  conducted a survey of 106 radio-loud AGNs, finding that about 20\% exhibit double-peaked Balmer lines. Of these, $\sim$ 60\% were well-fit by disk models. They noted that such "disk-like emitters" also tended to have unusually large equivalent widths in lines like [O\textsc{i}], [S\textsc{ii}], strong stellar contributions to the optical continuum, and very broad Balmer lines. \citet{Strateva2003} identified a larger population of double-peaked emitters from the Sloan Digital Sky Survey (SDSS), confirming that such profiles, although rare ( 3\% of AGNs), form a distinct class, mostly radio-loud.

While the accretion disk interpretation remains compelling for this type of objects, alternative explanations have been proposed. \citet{Gaskell1988} and \citet{Doan2020} suggested that double peaks could originate from binary black hole systems with dual broad-line regions. However, observational evidence, like the lack of expected radial velocity shifts over decades \citep[e.g.,][]{Halpern1988}, argues against this scenario for a fraction of objects.

Other alternatives include bipolar outflows or jets \citep{Zheng1990,Zheng1991}, which can mimic line asymmetries; disk plus wind models \citep{Popovic2004,Elitzur2014}, where a broader component arises from a wind or outflow overlaying the disk emission; non-axisymmetric disk structures, such as elliptical disks \citep{Eracleous1995} or spiral arms \citep{Lewis2010}, which can explain long-term variations and asymmetries in line profiles, and microlensing by stars in an intervening galaxy,  causing blue/red asymmetries \citep{Popovic2001}, especially in multiply-imaged quasars.

Polarization studies provided additional diagnostics for the emission-line geometry and disk signatures in prototypical objects like Arp 102B. \citet{Antonucci1996} predicted specific polarization behaviors for the H$\alpha$  based on a disk model, but found that observations did not fully match predictions. Nevertheless, they found evidence in favor of a disk origin, as indicated by the alignment between the polarization angles and the radio jet directions. Other observational efforts like those of \citet{Storchi2017} used Palomar survey data to show that broad Balmer line profiles in Seyfert 1 galaxies could often be decomposed into a disk-like component and a secondary Gaussian, suggesting that even in radio-quiet AGNs, a disk-like BLR structure may be common, though overlaid by emission from surrounding clouds or outflows.

In summary, the data collected over the past years indicate that a significant portion of double-peaked Balmer line emitters mainly originate from the outer regions of an accretion disk. However, several additional components; hot inner ion tori, spiral arms, transient hotspots, and possible outflows, are required to explain the full range of spectral and variability features observed in such objects. The accretion disk interpretation remains the most self-consistent and physically motivated framework for explaining double-peaked emission lines in radio-loud AGNs, also providing an opportunity to explore the dynamics and structure of the gas very close to the central black hole.

\citet{Lacerna2016} analyzed 89 isolated elliptical galaxies ($z < 0.08$) from the UNAM-KIAS catalog \citep{Hernandez2010}, comparing their properties with those of ellipticals in dense environments such as the Coma supercluster, identifying a small fraction ($\sim$4\%) of blue, star-forming isolated ellipticals. These galaxies are characterized by relatively low stellar masses ($7 \times 10^{9}$–$2 \times 10^{10}$ M$_\odot$), young light-weighted stellar ages ($<$1 Gyr), probably evidencing recent cold gas accretion driving their rejuvenation. These galaxies contrast with the $\sim$25\% fraction predicted by semi-analytic models \citep{Niemi2010}, highlighting the role of isolated environments in sustaining residual star formation.  

More recently, integral field spectroscopy surveys such as MaNGA \citep{Bundy2015} have uncovered additional elliptical galaxies showing AGN with broad, double-peaked Balmer emission lines \citep[][in prep.]{Cortes2022}, placing objects like UK~613 in a broader statistical context of disk-emitting AGN candidates.  

Here we present results from a dedicated spectroscopic follow-up of UK~613 at the Asiago Astrophysical Observatory and the Observatorio Astronómico Nacional San Pedro Mártir (OAN-SPM), combined with archival SDSS data spanning $\sim$15 years. UK~613 stands out as the only isolated blue elliptical in the UNAM-KIAS sample hosting a broad-line AGN with double-peaked Balmer emission, as first noted by \citet{Lacerna2016}. However, our new observations reveal that the double-peaked features have disappeared, suggesting a possible transition or ``turn-off'' in the accretion process that we investigate in this work.  

The outline of the present paper is as follows. Section \ref{sec:properties} provides context with relevant results for UK 613, summarizing key photometric and spectroscopic findings presented in \citet{Lacerna2016}, based on single-epoch observations from the SDSS survey. Section \ref{sec:asiago} presents new multi-epoch follow-up optical spectroscopic observations of UK 613 conducted at Asiago and San Pedro M\'artir Observatories. Section \ref{sec:comparison} compares the previous spectroscopic observations with the new multi-epoch data, discussing their main differences and implications for the properties of UK 613. Section \ref{sec:models} offers a more detailed analysis of the broad emission lines using physically motivated modeling like an accretion disk, bipolar outflow and the possibility of a tidal disruption event. Section \ref{ref:multiwavelenght} presents a multifrequency analysis of all available data for UK 613, including the reconstruction of its spectral energy distribution. Section \ref{LSS-Environment} explores the  environment of UK 613 to assess whether the galaxy lies within a cosmic void or wall, and discusses how its location might influence the presence of the active nucleus. Finally, Section \ref{Conclusions} concludes with our discussion and final remarks.

\section{Properties of the Blue Isolated Elliptical UNAM-KIAS 613}
\label{sec:properties}
\subsection{Selection Criteria}

UK~613 is a nearby galaxy at a redshift of $z = 0.02692$, included in the UNAM-KIAS catalogue of isolated galaxies \citep{Hernandez2010}. This catalogue was constructed from the DR4plus release of the New York University Value-Added Galaxy Catalog (NYU-VAGC; \citealt{Blanton2005}), itself based on the Sloan Digital Sky Survey (SDSS). The UNAM-KIAS sample was designed to identify galaxies evolving in low-density environments, thereby minimising the influence of interactions with similar-sized neighbouring systems. The isolation criteria consist of the following conditions:
\begin{enumerate}
    \item The magnitude gap ($\Delta m_r$) between the candidate galaxy and the neighbour is $\geq 2.5$ mag;
    \item The projected distance between the two is $\Delta d\geq 100R$, where $R$ is the seeing-corrected Petrosian radius of the candidate galaxy measured in the $i$-band; and
    \item The line-of-sight velocity separation $|\Delta v|\geq 1000\ km\ s^{-1}$.
\end{enumerate}

\subsection{Photometric Analysis} 

\citet{Lacerna2016} presented a photometric characterization of the four blue isolated elliptical galaxies identified in the UNAM-KIAS catalog of Isolated Galaxies \citep{Hernandez2010}, three of which also match the definition of blue faint isolated ellipticals by \citet{Niemi2010}. Using a careful 2D image decomposition with \texttt{galfit} \citep{Peng2002,Peng2011}, the structural properties of UK~613 were obtained. We briefly summarize their main results here. 

According to the two-phase scenario \citep{Oser2010,Johansson2012,Huang2013}, the inner/intermediate components of elliptical galaxies are built through dissipative processes (cold accretion or gas-rich mergers), while the outer component is assembled later via non-dissipative processes (minor, dry mergers). The $r$-band light distribution of UK~613 was best fit with three S\'ersic components. The inner and intermediate components showed low indices ($n<1$), while the outer component had a higher $n$, consistent with ex-situ processes, as also discussed by \citet{Hopkins2009}.  

For UK~613, the best-fit S\'ersic indices for the inner components are $n < 2.0$ with $r_{\rm eff} < 0.6$ kpc, smaller than the mean values ($n=3.2\pm2.1$, $r_{\rm eff}=0.7\pm0.4$ kpc) reported by \citet{Huang2013}. Within uncertainties, this suggests a more disky inner structure, consistent with rejuvenation by cold gas accretion or post-merger disk regeneration. However, as noted by previous authors \citep{George2015,George2023}, SDSS images have limited sensitivity for detecting subtle structures in early-type galaxies. Residual maps from the \texttt{galfit} analysis revealed only faint filamentary features, likely associated with dust in the central regions.  

A complementary analysis of elliptical isophotes showed no significant sloshing patterns, indicating that the central regions are dynamically relaxed. Finally, a ($g-i$) color map revealed a radial gradient with bluer colors toward the nucleus and an external blue ring-like feature, supporting the scenario of dissipative cold gas infall and recent central star formation.  

\subsection{Spectroscopic Analysis} 

\citet{Lacerna2016} also carried out a first spectroscopic study of UK~613 using SDSS spectra and stellar population synthesis with \texttt{starlight} \citep{Cid2005}, complemented with a PCA analysis following \citet{Hao2005}. Given the 3 arcsec SDSS fiber, both nuclear emission and stellar absorption had to be modeled simultaneously. UK~613 emerged as the only blue isolated elliptical in the UNAM-KIAS catalog hosting broad emission lines, thus classified as a Seyfert~1/LINER~1.  

The stellar population analysis encodes important clues about its assembly history. Mass-weighted star formation histories indicate that less than 5\% (20\%) of its stellar mass formed in the last 1 (3) Gyr, while light-weighted histories show that 30–60\% of its luminosity comes from stars younger than 1 Gyr. This suggests early mass assembly with a rejuvenation episode of star formation in the last $\sim$1 Gyr.  

The star formation timescale (SFTS; \citealt{Plauchu2012}), defined as the difference between mass- and light-weighted ages, was found to be $\sim$8.5 Gyr in UK~613. For comparison, early-type galaxies in compact groups show values around 3.3 Gyr, while isolated ellipticals have $\sim$5.4 Gyr. Thus, UK~613 formed its stars over a longer timescale, consistent with extended star formation activity.  

In summary, although UK~613 shows no strong morphological disturbances or merger signatures, its stellar populations reveal central rejuvenation within the last $\sim$1 Gyr, in parallel with the presence of a Type~I AGN. \citet{Lacerna2016} concluded that UK~613 assembled early like most ellipticals but was rejuvenated by recent cold gas accretion. The rarity of blue, star-forming isolated ellipticals in the UNAM-KIAS sample ($<4$\%) underscores the inefficiency of cold gas accretion from the intergalactic medium.  

A forthcoming environmental study of blue ellipticals with MaNGA IFS data (Hern\'andez-Toledo et al., in prep.) will provide crucial constraints on the formation and evolution of this rare class of galaxies.  

\subsection{\texttt{Starlight} Population Synthesis Models} 

\citet{Lacerna2016} applied \texttt{starlight} to the SDSS spectrum of UK~613 using 45 SSPs from \citet{Bruzual2003} with a Chabrier IMF. The SSP library included three metallicities (Z = 0.004, 0.02, 0.05) and 15 ages between 0.001 and 13 Gyr, with extinction modeled as a uniform dust screen \citep{Cardelli1989}. No power-law continuum was included in that fit.  

\begin{figure}
    \centering
    \includegraphics[width=\columnwidth]{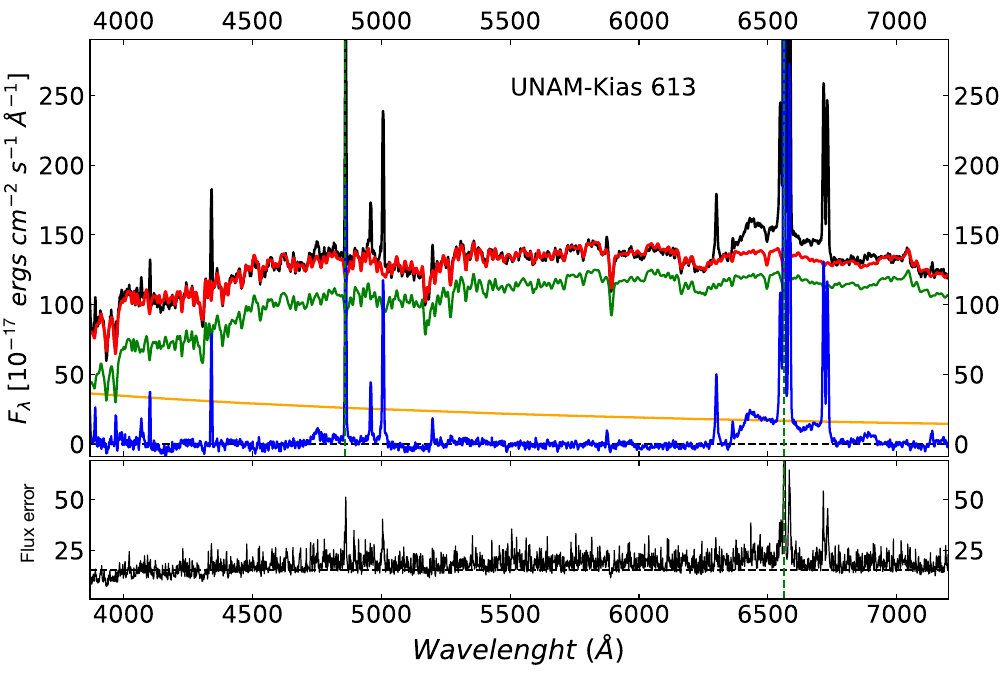}
    \caption{Top panel: UNAM Kias 613 spectrum from SDSS Data Release 7 (black line). A stellar population synthesis was performed with STARLIGHT obtaining the best model (red line), the host galaxy spectrum (green line), the AGN power law (orange line) and the emission spectrum obtained by subtracting the best model to the observed spectrum (blue line). Bottom panel: Flux errors of the model.}
    \label{fig:starlight}
\end{figure}

In this work, we performed a new \texttt{starlight} analysis of UK~613 with an extended library of 150 SSPs from \citet{Bruzual2003}, covering six metallicities (Z = 0.0001–0.05) and 25 ages (0.001–13 Gyr). As before, extinction was modeled with the \citet{Cardelli1989} law. In addition, we included five power-law components to represent a possible AGN continuum. Figure~\ref{fig:starlight} shows the best-fitting model (red), which combines the host galaxy spectrum (green) with a power-law component (orange). The resulting emission-line spectrum, obtained after subtracting the stellar model, clearly reveals multiple broad components around H$\alpha$ $\lambda$6563\,\AA\ (blue).

\section{New Spectroscopic Observations of UNAM-KIAS 613}
\subsection{Asiago Observations}
\label{sec:asiago}

A new spectroscopic dataset for UNAM-KIAS 613 was acquired in December 2018 at the Asiago Astrophysical Observatory using the 1.82-meter Copernico Reflector Telescope. Three individual exposures with 1.69 arcsec long slit, each of 1800 seconds, were obtained. The observations covered a broad wavelength interval, encompassing both the H$\alpha$ and H$\beta$ emission line regions, as well as the intervening continuum. However, this broad spectral range came at the cost of lower spectral resolution, which prevented reliable deblending of the narrow [N\,\textsc{ii}] $\lambda\lambda$6548,6584 doublet from the H$\alpha$ line.

\begin{figure}
    \centering
    \includegraphics[width=\columnwidth]{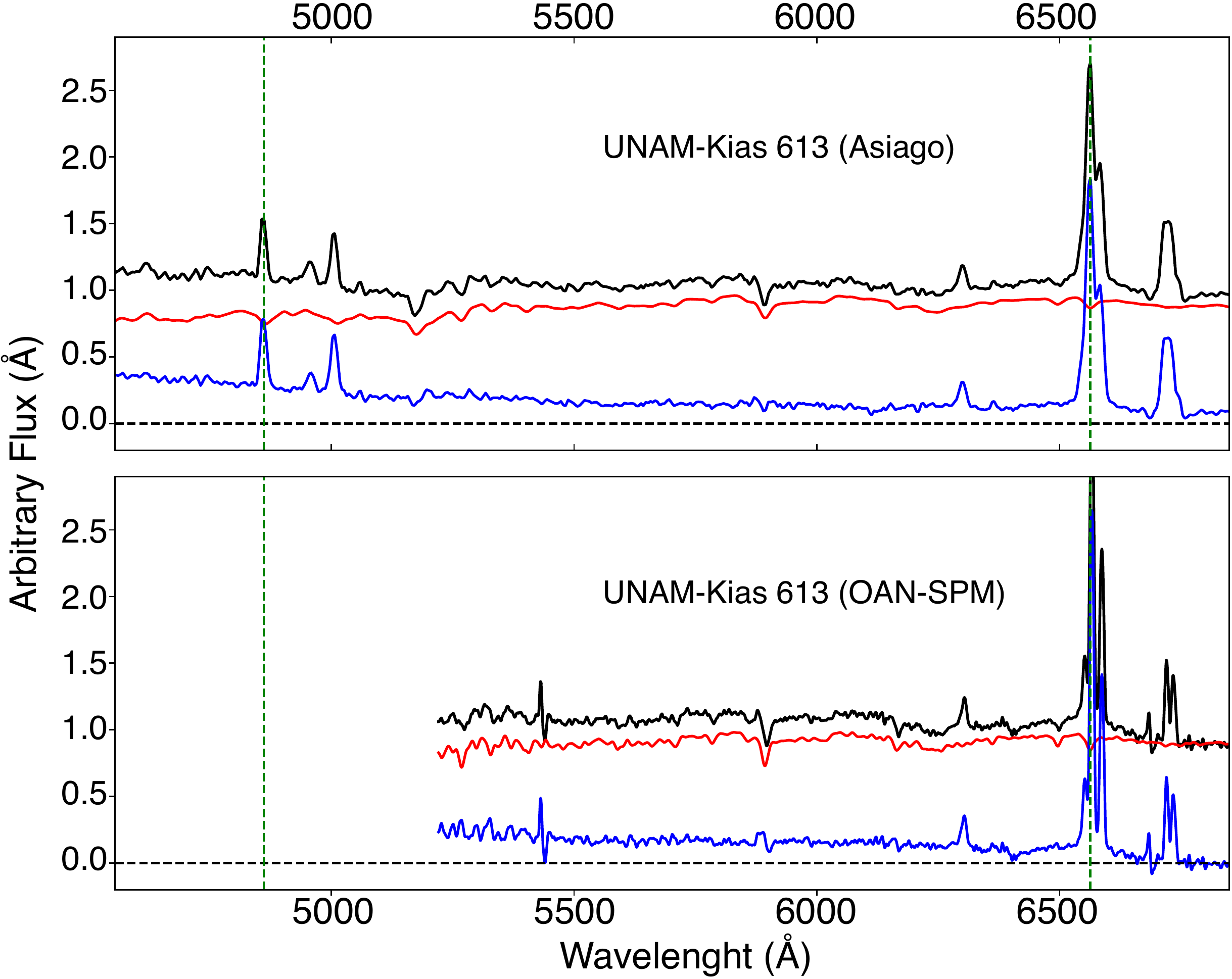}
    \caption{Asiago spectrum (top panel) and OAN-SPM spectrum (bottom panel) for UNAM-KIAS 613. In both cases, the stellar population model derived from the SDSS spectrum (red line) was subtracted to isolate the AGN emission-line spectrum (blue line). The observed spectra are shown in black.}
    \label{fig:asiago}
\end{figure}

Data reduction was carried out using standard long-slit procedures in the \textsc{IRAF} software package \citep{Tody1986}. The final reduced spectrum is shown in the top panel of Figure~\ref{fig:asiago}, where the black line represents the observed spectrum. 
A first visual inspection reveals that the broad shoulders previously present in the H$\alpha$ region in the 2006 SDSS spectrum are no longer visible, suggesting a possible change in nuclear activity over the 12 years that followed.

\subsection{San Pedro M\'artir Observations}
\label{sec:spm}

To confirm the possible long-term variability in the nuclear spectral features of UNAM-KIAS 613, a third spectroscopic dataset was obtained in April 2023 at the National Optical Observatory of San Pedro M\'artir (OAN-SPM), using the 2.12-meter Reflector Telescope. Three individual exposures with $\sim$ 2 arcsec longslit of 1200 seconds each were acquired, targeting the blue (H$\beta$) and red (H$\alpha$) spectral regions separately. In contrast to the Asiago observations, these data were obtained at higher spectral resolution, allowing for better separation of individual narrow emission lines.

Data reduction was performed using a custom pipeline based on the \texttt{PyKOSMOS} package \citep{davenport2021}, which relies on the \texttt{astropy} library for its core functionality. The pipeline automates standard reduction steps, including bias subtraction, flat-field correction, and sky subtraction. For sky removal, a region of the two-dimensional spectrum free of target emission was selected to extract an average sky spectrum, which was subtracted from the science frame. Wavelength and flux calibration were performed using standard arc lamps and the spectrophotometric standard star Feige 34, for which we adopted a high-resolution model from the ESO library \citep{Oke1990}.

The final wavelength- and flux-calibrated spectrum is shown in the bottom panel of Figure~\ref{fig:asiago}. As with the Asiago spectrum, the absence of broad components in the wings of H$\alpha$, previously observed in 2006, is confirmed, supporting the hypothesis of a decrease in nuclear activity and/or the disappearance of transient outflow features over the last 17 years.

\section{Spectroscopic Results and AGN Properties}
\label{sec:comparison}
\subsection{Spectral Evolution of UNAM-KIAS 613}

Our compilation of spectroscopic observations for UNAM-KIAS 613 reveals clear spectral evolution over a time span of 17 years, from the original SDSS data obtained in 2006 to the most recent OAN-SPM observations in 2023. This extended temporal baseline provides a valuable opportunity to investigate long-term changes in the nuclear emission features of the galaxy.

All observations were centered on the nucleus. The SDSS spectrum was obtained using a 3'' diameter circular fiber, while the follow-up observations employed long-slit spectroscopy. The OAN-SPM data were acquired with a slit width of 2'' and extracted over 20 pixels along the spatial direction, corresponding to $\sim$11.8'' for a spatial scale of 0.59'' per pixel, resulting in an effective aperture of $\sim$2'' $\times$ 11.8''. Similarly, the Asiago spectrum was obtained with a slit width of 1.69'' and extracted over 30 pixels ($\sim$7.5'' for a scale of 0.25'' per pixel), yielding an aperture of $\sim$1.69'' $\times$ 7.5''. These correspond to effective areas of $\sim$7.1 arcsec$^2$ (SDSS), $\sim$12.7 arcsec$^2$ (Asiago), and $\sim$23.6 arcsec$^2$ (OAN-SPM). Although the aperture geometries and sizes differ, they are all much larger than the expected size of the BLR ($\lesssim$0.01 pc), ensuring that the broad-line emission is fully enclosed in all cases and that its kinematic properties can be directly compared. While these differences may introduce varying contributions from extended emission (e.g., from the narrow-line region and host galaxy) and affect the relative fluxes of narrow lines, they do not impact the morphology of the broad H$\alpha$ components analyzed in this work. Differences in host-galaxy contamination were minimized by modeling and subtracting the stellar continuum using \texttt{starlight} for all epochs.

To enable a direct comparison between the SDSS, Asiago, and OAN-SPM spectra, all datasets were normalized using the integrated flux of the [N\,\textsc{ii}] $\lambda\lambda$6717,6731\,\AA\ + H$\alpha$ narrow component. Additionally, the SDSS and OAN-SPM spectra were convolved to match the instrumental resolution of the Asiago data. Figures~\ref{fig:starlight} and~\ref{fig:asiago} illustrate that the SDSS H$\alpha$ profile presents a visibly more complex morphology (pronounced shoulders and extended wings) than the later Asiago and OAN-SPM spectra. The similarity between the Asiago and OAN-SPM profiles, together with the disappearance of the SDSS shoulders, suggests a change in the BLR kinematics (or a transient event) between epochs. The time intervals between the SDSS and Asiago observations, and between Asiago and SPM, are approximately 12 and 5 years, respectively. To explore these spectral differences in detail, we performed a multi-component emission line fitting analysis, which is described in the next section.

Since the stellar population of the host galaxy can dominate the optical continuum and potentially mask the broad emission-line components, and given that it is expected to remain unchanged over the timescale considered, we subtracted the stellar continuum model derived from the SDSS spectrum from both the Asiago and OAN-SPM observations to isolate the AGN emission. Prior to subtraction, the model was convolved to match the spectral resolution of each dataset. The results of this process are presented in Figure~\ref{fig:asiago}, where the stellar model is shown in red and the resulting emission-line spectrum—after subtraction—is displayed in blue.

\subsection{Emission lines fitting} 
\label{specfit}

We used the \texttt{specfit} routine \citep{Kriss1994} within the \texttt{iraf} environment to fit the emission-line spectra obtained after stellar continuum subtraction. This empirical fitting approach, which includes broad-line components, serves as a useful first approximation for characterizing the properties of the central supermassive black hole using indirect diagnostics.

The fits were performed in the H$\alpha$ spectral region (6200–6800\,\AA), following the methodology outlined in \citet{Negrete2018}, with several adaptations. The model includes a power-law continuum, narrow and broad emission-line components, and optical Fe\,\textsc{ii} emission. Initial estimates were provided for all model components. Based on simulations of AGN spectra by \citet{Marziani2003}, the minimum detectable Fe\,\textsc{ii} emission in the optical was visually calibrated as a function of H$\beta$ line width and signal-to-noise ratio. 

All spectra were corrected to the rest-frame prior to fitting. The centroids of the narrow lines were fixed to their rest-frame wavelengths, assuming that they originate in the narrow-line region (NLR). In contrast, the centroid of the broad H$\alpha$ component was left as a free parameter, consistent with the expectation that broad-line emission arises from a distinct, more dynamically active region.

The width of the narrow H$\alpha$ line was first determined and then used to constrain the widths of the other narrow components. For forbidden doublets such as [O\,\textsc{iii}] and [N\,\textsc{ii}], both the line separation and theoretical flux ratios were fixed. For the [S\,\textsc{ii}] doublet, only the separation between components was constrained. In contrast, the full width at half maximum (FWHM) of the broad H$\alpha$ components was left entirely free, reflecting their origin in the broad-line region (BLR).

Each narrow emission line was modeled with a single Gaussian component representing the NLR emission. We also tested the inclusion of an optional secondary narrow gaussian to explore the presence of a kinematically distinct narrow-line system (as observed in some dual or offset AGN), but this additional component was not required by the data. The broad H$\alpha$ profile was modeled using a combination of three gaussians, following the approach of \citet{Lacerna2016} (see also \citealt{Lewis2010, Storchi2017}).

The fitting procedure employed the Levenberg-Marquardt algorithm for non-linear least-squares minimization, optimizing the fit to the observed flux as a function of wavelength. If the solution failed to converge or did not reproduce the observed profile adequately, the initial parameters were adjusted manually and the fitting process repeated. The goodness of fit was evaluated using the reduced chi-squared statistic ($\chi^2_{\nu}$), with values close to unity considered indicative of a satisfactory fit.

\begin{figure*}
    \centering
    \includegraphics[width=2.1\columnwidth]{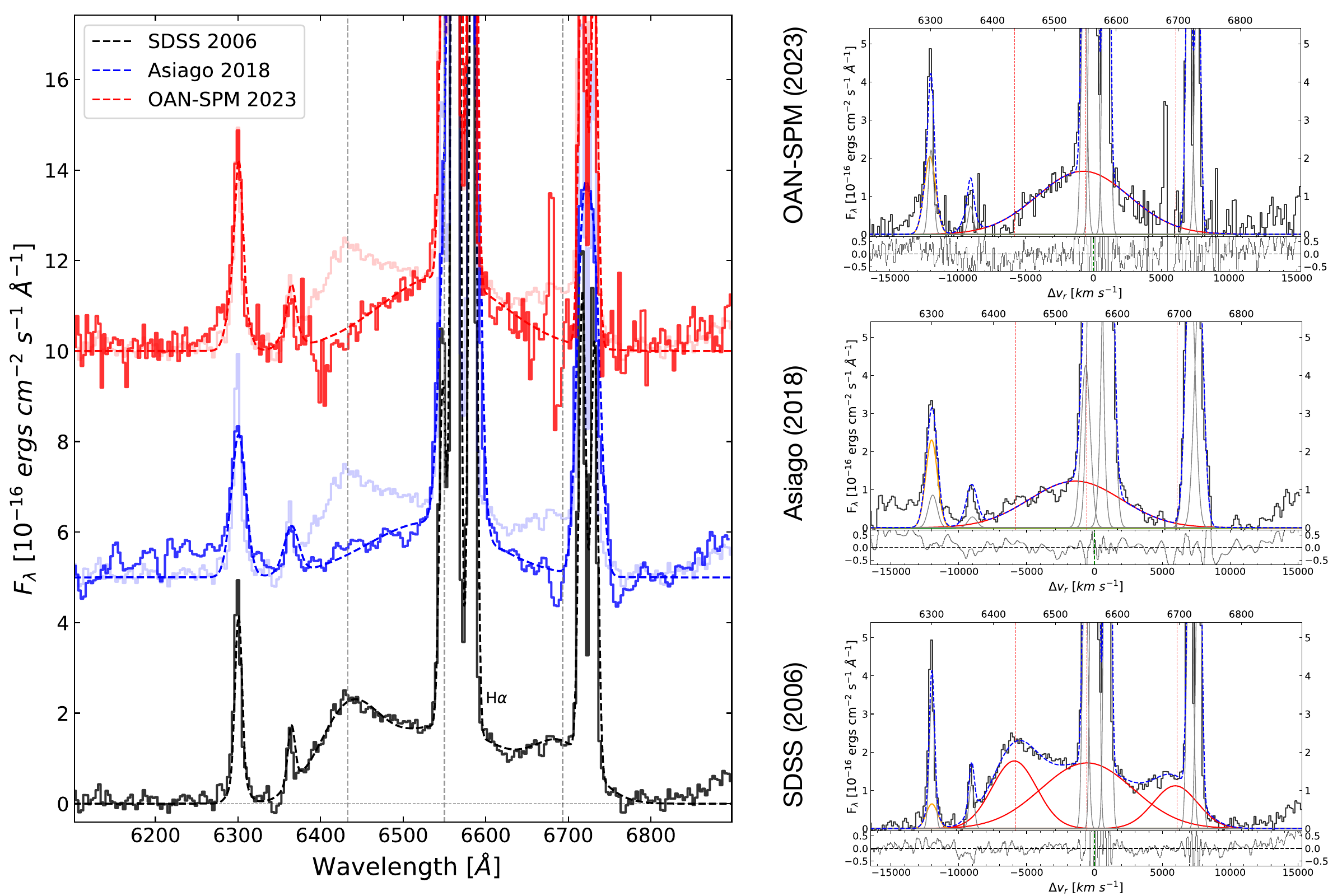}
    \caption{Left figure: Comparison between the H$\alpha$ region after normalizing by the [N\textsc{ii}]+H$\alpha$ flux. The SDSS spectrum is shown in all spectra. Right figures: The emission lines fitting procedure of each observation. In black, the observed spectrum, blue is the best model, gray is the narrow emission lines, and red is the broad emission lines.}
    \label{fig:haplotSDSS}
\end{figure*}

Table~\ref{tab:NarrowProperties} presents the measured fluxes and FWHM values of the narrow emission lines, along with the continuum flux in the 6200–6250\,\AA\ range. The right panels of Figure~\ref{fig:haplotSDSS} show the results of our emission line decomposition for the SDSS, Asiago, and OAN-SPM spectra. In each panel, the observed spectrum is shown as a black line, the best-fit model as a segmented blue line, the broad components in red, and the narrow components in gray. The left panels provide a direct comparison of the H$\alpha$ region.

To highlight the changes in the H$\alpha$ profile across epochs, the Asiago and OAN-SPM spectra are overlaid with the SDSS spectrum in the background. For this comparison, all spectra were normalized using the combined flux of [N\,\textsc{ii}] $\lambda\lambda$6548,6583 and the narrow H$\alpha$ component. The normalization process is detailed in Table~\ref{tab:normalization}, which reports the flux ratios between the continuum at 6200–6250\,\AA\ and key narrow emission lines: [O\,\textsc{i}] $\lambda$6300, [N\,\textsc{ii}]+H$\alpha_{\rm NC}$, and [S\,\textsc{ii}] $\lambda\lambda$6717,6731. These ratios allow us to assess flux differences across the spectra. To scale all spectra to a common reference level, we used the [N\,\textsc{ii}]+H$\alpha_{\rm NC}$ flux from the SDSS observation. The last column of Table~\ref{tab:normalization} shows the scaled continuum flux values in the 6200–6250\,\AA\ window, which indicate an increase in continuum level in the most recent SPM spectrum.

\begin{table}
    \centering
    
    \begin{tabular}{cccccc}
    \hline
    \hline
     & & & [O\textsc{i}]  & [N\textsc{ii}]+H$\rm\alpha_{NC}$ & [S\textsc{ii}]\\
    Spectrum & $f_{6200-6250}$ & FWHM & flux  & flux & flux  \\
    \hline
    SDSS & 1.657 & 368 & 51 & 770 & 212 \\
    Asiago & 0.763 & - & 27 & 293 & 98 \\
    SPM & 2.681 & 384 & 83 & 901 & 230 \\      
    \hline
    \end{tabular}
    \caption{Main properties of UK 613 obtained with the emission lines fitting procedure. Second column is the continuum flux at 6200-6250\AA, third column the FWHM of the narrow emission lines, fourth column the [O\textsc{i}]$\rm\lambda$6300 emission line flux, fifth column the combined flux of [N\textsc{ii}]$\rm\lambda\lambda$6548,83 and H$\rm\alpha_{NC}$, and sixth column the combined flux of [S\textsc{ii}]$\rm\lambda\lambda$6717,31. Flux units are 10E-16 erg/cm$^2$/s/\AA}
    \label{tab:NarrowProperties}
\end{table}

As illustrated in Figure~A.3 of \citet{Lacerna2016}, Figure~\ref{fig:haplotSDSS} clearly shows three distinct broad components in the H$\alpha$ profile of the SDSS spectrum, with the central component being the broadest. Vertical dashed lines indicate the centroids of these components. The blue- and redshifted components flanking the central peak suggest complex kinematics, possibly linked to transient phenomena such as bipolar outflows, accretion disk instabilities, or a tidal disruption event. In contrast, the Asiago and OAN-SPM spectra only show the central broad component, supporting the hypothesis that the additional components seen in 2006 were transient.

The measured FWHM of the central broad component in the SDSS spectrum is $\sim$8025\,$\pm$\,25\,km\,s$^{-1}$, while the red and blue components have FWHM values of $\sim$3750 and $\sim$3950\,km\,s$^{-1}$, respectively. Despite the disappearance of the side components in recent observations, the central component remains, and its FWHM and luminosity are consistent across epochs. The luminosity of the broad H$\alpha$ component varies between $\log L_{\rm H\alpha} \sim 40.6$ and $40.75$, depending on the adopted normalization—whether using [N\,\textsc{ii}]+H$\alpha$ or [O\,\textsc{i}].

UK 613 fits as a low-luminosity Seyfert 1 AGN, edging toward extreme Population B \citep{sulenticetal00a,marzianietal22b}, where Population B sources are characterized by broad Balmer lines (FWHM H$\beta \gtrsim 4000 $ km/s) weak $R_\mathrm{FeII}$, and  $\lambda_{\rm Edd}$\ close to  the Eddington ratio lower limit observed in the $M_\mathrm{BH}$ - luminosity diagram \citep{marzianietal03b}.  The large FWHM $\sim$ 16500 km/s measured on the accretion disk profile and the absence of FeII emission in the AGN spectrum places the source  in spectral type B1+, close to the B1++ lower limit. This is the region of the quasar main where low accretors are typically located, although in the samples that were used to define the MS in the original papers by \citet{borosongreen92} and \citet{sulenticetal00a} sources as faint as UNAM-KIAS 613 were not considered.  

\subsection{Estimates of accretion parameters}

We estimate the mass of the supermassive black hole mass $M_\mathrm{BH}$\ in UNAM-KIAS 613 using the single-epoch virial method of \citet{Greene2005}, which relates the black hole mass to the FWHM and luminosity of the broad H$\alpha$ emission. We base the estimate on   the central broad H$\alpha$\ component in the OAN-SPM spectrum. For consistency, we apply this method only to the central broad component identified in both the SDSS and SPM spectra. In Sections \ref{disk} and \ref{bicone} we further motivate the choice of selecting the central component only.

\begin{table*}
    \centering
    \begin{tabular}{cccccccc}
    \hline
    \hline
    Spectrum   &   \multicolumn{3}{c}{Normalizations} & Scaled Continuum \\ \cline{2-4}
     & [O\textsc{i}] & H$\rm \alpha_{NC}$ + [N\textsc{ii}]  & [S\textsc{ii}] & $f_{6200-6250}$\\
    \hline

    $\lambda f_\lambda$ SDSS & 0.0325 & 0.0022 & 0.0078  & 1.658$\pm$0.231\\
    $\lambda f_\lambda$ Asiago & 0.0287 & 0.0026 & 0.0078 & 2.005$\pm$0.271 \\
    $\lambda f_\lambda$ SPM & 0.0323 & 0.0030 & 0.0116 & 2.292$\pm$0.330 \\      
    \hline
     $\lambda f_\lambda$ Asiago/$\lambda f_\lambda$ SDSS & 0.88 & 1.23 & 1 \\
     $\lambda f_\lambda$ SPM/$\lambda f_\lambda$ SDSS & 0.99 & 1.38 & 1.49\\
    \hline
    \end{tabular}
    \caption{Continuum variability assessment with different normalizations in the H$\alpha$\ spectral range. Flux units are 10E-16 erg/cm$^2$/s/\AA}.
    \label{tab:normalization}
\end{table*}

Additionally, we estimate the SMBH mass using the single-epoch virial scaling relations of \citet{Vestergaard2006}, which are based on the FWHM of broad H$\beta$ and the continuum luminosity at 5100\,\AA. Since broad H$\beta$ is not detected in any of the spectra, we infer its FWHM from the measured FWHM of H$\alpha$ using the empirical relationship between the two broad Balmer lines derived by \citet{Shen2008}, namely  FWHM(H$\beta$) /FWHM(H$\alpha$) $\approx$ 1.16 $\pm$ 0.02.

To estimate the 5100\,\AA\ continuum luminosity, we assume a power-law continuum of the form $L_{\lambda} \propto \lambda^{\alpha}$, with a typical AGN slope of $\alpha = -1.5$    \  \citep{marzianietal03b}.   Under this assumption, we derive $L_{5100} \approx 1.35 \times L_{6200}$. The resulting empirical scaling factor  between the 5100\,\AA\ and 6200–6250\,\AA\ luminosities from the SDSS spectrum is of $\approx$1.28. Given the consistency of the host galaxy and the AGN spectral shape across epochs, we adopt this empirical factor for all three spectra. 

Table~\ref{tab:broadproperties} summarizes the $M_\mathrm{BH}$\ mass estimates derived from both methods, alongside the corresponding 5100\,\AA\ luminosities, FWHMs, and Eddington ratios ($\lambda_{\rm Edd}$) for the   broad H$\alpha$ and H$\beta$ components. While the masses obtained using the \citet{Vestergaard2006} relation are systematically higher, the \citet{Greene2005} method provides a more direct and reliable estimate in this context. The  SMBH masses resulting from utilizing the H$\alpha$ FWHM lie in the range $\log M \sim 7.18$–7.26 [M$_{\odot}$] when considering only the central broad component, and increase up to $\sim$7.58 when all three broad components of H$\alpha$ in the SDSS spectrum are included. The  H$\alpha$\ mass estimates yield Eddington ratios ranging from $\lambda_{\rm Edd} \approx 0.029$ to $0.040$. This places UK 613 in the regime of sub-Eddington accretion, typical of low-luminosity AGNs. This estimate   suggests that UK 613 is close to the boundary between radiatively efficient and   inefficient accretion. 

\begin{table*}
    \centering
    \begin{tabular}{ccccccccc}
    \hline
    \hline
     Spectrum & logL$\rm_{bol}$5100 & logL$\rm_{H\alpha}$  & FWHM($\rm{H\alpha}$) & FWHM($\rm{H\beta}$) & logM$\rm_{BH}^{(a)}$ & $\rm\lambda_E^{(a)}$ & logM$\rm_{BH}^{(b)}$ & $\rm\lambda_E^{(b)}$\\
    \hline
    SDSS & 43.82  & 40.75 (41.01) & 8000 (9679) & \textit{9089 (10975)} & \textit{7.97 (8.13)} & \textit{0.006 (0.004)} & 7.26 (7.58) & 0.029 (0.014) \\
    Asiago & \textit{43.89} & 40.60 & 8048 & \textit{9143} & \textit{8.01} & \textit{0.006} & 7.18 & 0.040 \\
    OAN-SPM & \textit{43.94} & 40.73 & 8000 & \textit{9089} & \textit{8.04} & \textit{0.006} & 7.25 & 0.038 \\
    \hline
    \end{tabular}
    \caption[Measurement of physical parameters]{Physical parameters measurement and estimations from the emission lines fitting procedure. $\rm^{(a)}$ Estimated from \cite{Vestergaard2006}. $\rm^{(b)}$ Estimated from \cite{Greene2005}. Italic values are indirect values. For the SDSS spectrum, we estimate the parameters using the three broad components of H$\alpha$ shown in parenthesis.}

    \label{tab:broadproperties}
\end{table*}

\section{Interpretation of the line profile}
\label{sec:models}

An intriguing aspect of UNAM-KIAS 613 is the disappearance of its blue and red broad H$\alpha$ components observed in the SDSS spectrum (2006), which are no longer visible in the subsequent spectra from Asiago (2018) and OAN-SPM (2023). This spectral evolution over 17 years raises the question of whether similar transient broad line behavior has been documented in other active galactic nuclei.

\citet{Marsango2024} presented a long-term spectroscopic monitoring of Pictor~A, a well-known double-peaked emitter. Early observations in the 1990s detected a change from an apparently single-peaked to a clearly double-peaked H$\alpha$  profile \citep{Halpern1994,sulenticetal95}.  Over 2006–2018, its H$\alpha$ profile displayed substantial variability in peak intensity and structure, yet the characteristic red and blue peaks never disappeared entirely, and the underlying disk-like morphology remained visible at all epochs. Similarly, most classical double-peaked emitters—such as Arp~102B, NGC~1097, and NGC~7213—show changes in peak ratio, separation, or asymmetry, but the two-peak structure persists even under strong variability \citep[e.g.][]{Shapovalova2013,Storchi1997,Eracleous2003,Schimoia2015,Schimoia2017}. Even in changing-look AGN, where the broad-line region undergoes dramatic luminosity changes, disk-like double-peaked profiles typically reappear in rms spectra \citep[e.g.][]{Popovic2023}. In contrast to these cases, UNAM-KIAS~613 is remarkable in that both disk-like peaks seen in 2006 disappear completely in later epochs, leaving only a single broad component with no detectable shoulders. Such a full suppression of double-peaked features is rarely documented among known double-peaked AGN, underscoring the unusual nature of its broad-line evolution.

The seyfert 2 galaxy NGC2110, in which extremely broad, double-peaked H$\alpha$ emission was detected only in polarized flux; subsequent direct spectra from the Hubble STIS instrument in 2000 lacked this feature, indicating that the double-peaked component vanished in direct view \citep{Moran2007, Tran2010}. This suggests that the feature was either obscured or ephemeral, possibly tied to scattering clouds or transient BLR geometry.

In a statistical study of variable double-peaked AGN, \citet{Lewis2010} analyzed over 20 such sources and found that while peak shapes and fluxes varied significantly over multi-year timescales, the double-peaked profiles remained recognizable. These variations were attributed to changing emissivity patterns in the disk rather than to episodic phenomena like outflows or tidal events.

In contrast, UNAM-KIAS 613 presents a more extreme behavior. It suggests a transient event that produced a temporary outflow or a kinematically distinct structure, possibly linked to a short-lived episode of AGN activity. The nature of this transient component, perhaps a bipolar wind or a disrupted disk structure, requires further investigation.

\subsection{The disk model for UK 613}
\label{disk}

In addition to the empirical fitting of the broad emission-line components described in the previous section, we also explored the possibility of modeling the H$\alpha$ profile using an accretion disk (AD) model, motivated by the distinctive red and blue broad components observed in the SDSS spectrum of UNAM-KIAS 613. The overall shape of the H$\alpha$ profile—characterized by a broad base, prominent peak separation, and asymmetric intensity between the blue and red sides—closely resembles that of the prototypical AD emitter Arp 102B \citep{Chen1989,Eracleous1995,storchi-bergmannetal93,Storchi2017}.

\begin{figure}
    \centering
    \includegraphics[width=\columnwidth]{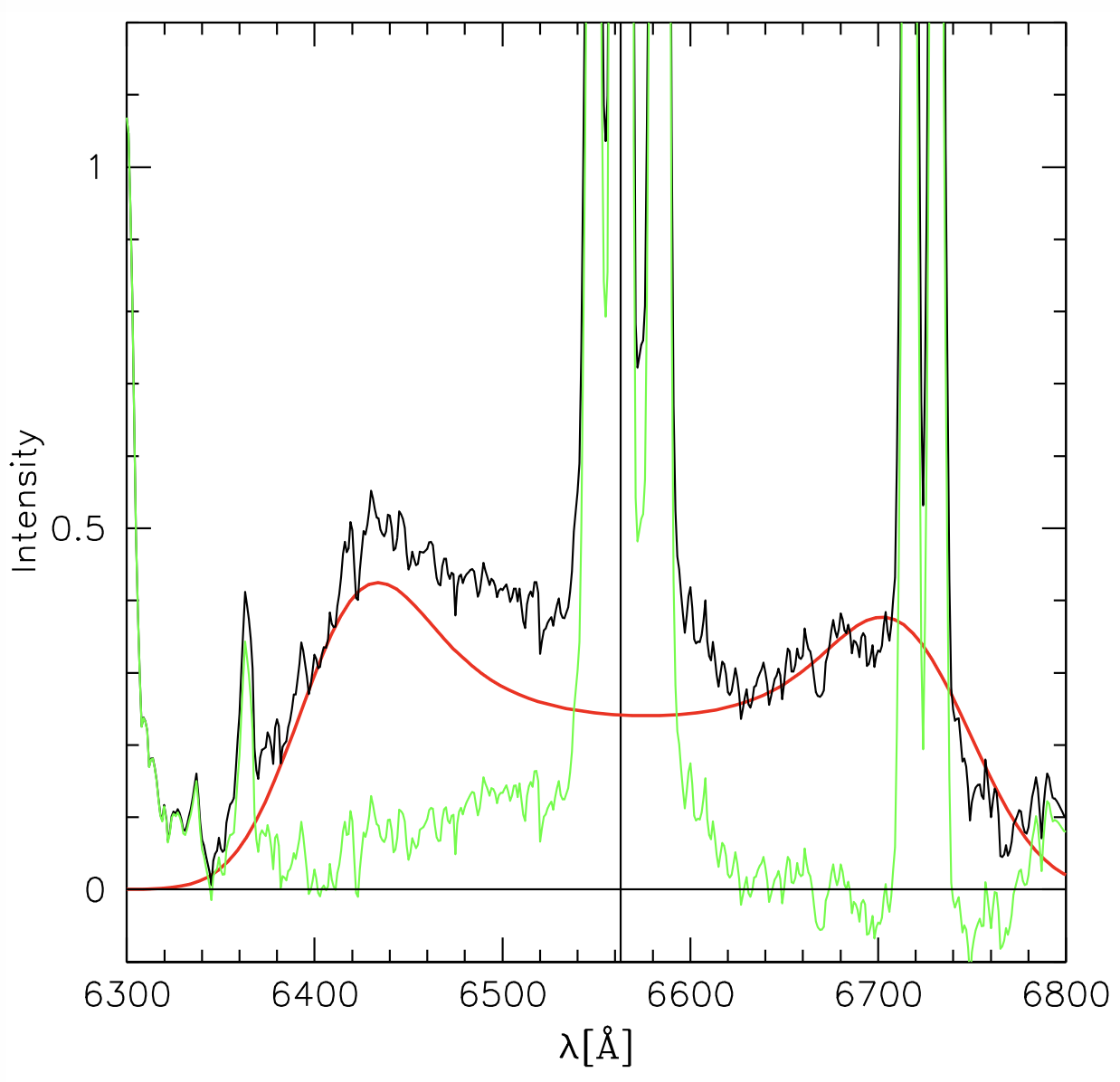}
    \caption{The disk model of UNAM-Kias 613 following \citet{Chen1989} (red line). The black line traces the continuum substract H$\alpha$ profile of the SDSS spectrum; the green line the residual after the subtraction of the disk model.}
    \label{fig:diskmodel}
\end{figure}

Following the formalism of \citet{Chen1989}, we applied a relativistic disk profile model to the continuum-subtracted SDSS spectrum. Although no full formal fit was performed—owing to the presence of excess emission superimposed on the disk profile that is difficult to parameterize—the model successfully reproduces the main features of the line: the large width, redward asymmetry, and the slopes of the wings. Figure~\ref{fig:diskmodel} shows the resulting best-match disk profile, obtained using the following parameters: inner radius $r_\mathrm{in} = 950\,r_\mathrm{g}$, outer radius $r_\mathrm{out} = 2000\,r_\mathrm{g}$, inclination angle $\theta \approx 66^{\circ}$, emissivity exponent $a = 2$ (corresponding to an emissivity law $\epsilon(r) \propto r^{-2}$), and local line broadening of $\sim 4000$\,km\,s$^{-1}$. We note, however, that this inclination is relatively high for the assumptions of the original \citet{Chen1989} modeling framework, and caution should be taken in interpreting the results quantitatively.

The blue peak appears stronger than the red, and the broad separation between them suggests that the emission arises from a disk viewed at significant inclination, with a relatively small ratio of outer to inner radius. Residual broad emission above the disk model has been observed in other well-known disk emitters, such as Arp 102B \citep{Chen1989}. The H$\alpha$ profile in its entirety might be associated with a non-axisymmetric structure, such an elliptical disk \citep{Eracleous1995,Storchi1997}. If we interpret the residual emission with respect to the \citep{Chen1989} profile shown in Fig. \ref{fig:diskmodel} as an excess from an additional component, the excess might originate from  elliptical secondary rings \citep[e.g.,][]{Bon2018}, or may trace other unresolved kinematic structures like outflows \citep{2012ApJ...753..133F}, or non-axisymmetric short-lived emitting regions such as hot spots \citep{flohicetal08,jovanovicetal10,popovicetal14}. While more complex disk geometries could improve the fit, the current model captures the essential characteristics of an inclined accretion disk as the dominant source of the observed line profile.

Assuming that the disk model is representative of the dynamics of the broad line emitting region, an additional $M_\mathrm{BH}$ estimate can be based directly on the virial black hole mass equation.  The virial mass equation is $M_{\rm BH} = \frac{r \delta v_{\rm K}^{2}}{G} = f_{\rm S}(\theta) \frac{r  {\rm FWHM}^{2}}{G}$. The structure factor $f_\mathrm{S}(\theta)$ connects the observed line broadening measured by the line FWHM to the Keplerian velocity $v_{\rm K}$, and  is given as a function of the viewing angle $f_{\rm S}(\theta)  = \frac{1}{4 \left[ \frac{1}{3}\left(\frac{\delta v_{\rm iso}}{\delta v_{\rm K}}\right)^{2} + \sin^{2} \theta  \right]   } \label{eq:fs}$ \citep{jarvismclure06,decarlietal11},  and becomes $\approx $ 0.3, for a highly flattened distribution $\frac{\delta v_{\rm iso}}{\delta v_{\rm K}}\sim 0.1$ observed at high inclination ($\theta \approx 66$). Here we are assuming the the velocity field is composed of a Keplerian component aligned with the BH equatorial plane, and an isotropic component $\delta v_{\rm iso} \ll \delta v_{\rm K}$. The FWHM of H$\alpha$ of the SDSS spectrum is $\approx 14600$ \kms. Assuming the BLR radius radius $r$ as obtained from the correlation with luminosity $\log r \approx (1.527) +0.533 \cdot \log [\lambda L_\lambda(5100)/10^{44}]$  ld \citep{bentzetal13}, the black hole mass is   $\log M \sim 6.82$ [M$_{\odot}$], close (perhaps fortuitously) to the values obtained from the scaling laws of \citet{Greene2005}.  At any rate,  both methods consistently indicate  that the source is  a relatively low mass system accreting at a very low rate, close to a critical lower limit in L/L$_{\rm Edd}$ for the radiatively efficient domain. The bolometric luminosity $L \sim 10^{43}$ erg s$^{-1}$ is well-above the minimum luminosity needed to maintain a virialized BLR \citep{Laor2003}. 

If the mass supply shuts off abruptly, the  draining time of a disk is set mainly by the viscous time toward the outer edge i.e., $
t_{\rm surv}\sim t_{\nu}(R_{\rm out}) \sim {R_{\rm out}}/{|v_\mathrm{r,drift}|}$,
and in an $\alpha$-disk the inward drift speed is of order $
|v_\mathrm{r,drift}|\sim \alpha \left({H}/{R}\right)^2 v_\mathrm{Kepl}$, implying that $
t_{\nu}(R)\sim {1}/{\alpha}\left({R}/{H}\right)^2 {1}/{\Omega_K}
$, with $\Omega_K=\sqrt{GM/R^3}$. Scaling to the parameters suitable for UK 613, we have:

$t_{\rm drift} 
\approx 2.8\times10^3\ 
M_7^{-1/2} R_{15}^{3/2}
\alpha_{0.1}^{-1}
({H/R})_{0.01}^{-2}\, \mathrm{yr}$

, where $R \sim 10^{15}$\ cm corresponds to $\sim 700$\ gravitational radii for $M_\mathrm{BH}$ $\sim 2 \cdot 10^7$ M$_{\odot}$. If the outer radius is at 2000 gravitational radii as indicated by the best fit (Fig. \ref{fig:diskmodel}), $t_{\rm drift} $\ becomes $\sim 10^4$ yr.     The implication is that an illuminated $\alpha$-disk should be relatively stable, and possibly fade over a relatively long timescale, much longer than the $\sim 10$ yr separating the SDSS from the Asiago spectrum.  

Only a few double peakers suffered a complete disappearance or appearance of the double-peaked profiles. The set of known double peakers should have a  disappearance/appearance rate $\lambda \sim n / (N \cdot \delta t) $, where $N$ is the number of double peakers pooled, $\sim 100$, $n$ the number of convincing cases of single-double transition (and vice versa, and of full disappearance) which we assumed to be $\approx 5 \lesssim 10$.  The survival time of the disk could therefore  be $t_{\rm surv} \sim 1/\lambda \sim 800 N_{100} \delta t_{40} / n_{5}$ yr. This number is highly uncertain, but shorter by an order of magnitude than the $t_{\rm drift}$. "Disk disappearance" could be a circumstance that is especially likely for double peakers.  After the first double-peaked profiles were identified in the 1980s \citep{Halpern1988}, double peakers were discovered as especially common at low accretion rates in hard-X-ray and radio-loud AGN, reinforcing that this is a real, low-accreting, sub-population of AGN rather than isolated outliers \citep{Eracleous1994,Strateva2003,marzianietal22b,wardetal25}. 

There might be other factors causing the disappearance of the double-peaked profiles and hence shortening their lifetime, other than the gas exhaustion; for example, obscuration may produce alteration of the double-peaked profiles \citep[][these authors provide several examples]{gaskellharrington18}, although it would need to be extremely fine-tuned to produce disappearance of the double-peaked profile, and should be  a short-lived episode  \citep{kaastraetal14}.

A more likely possibility is a change in geometry or optical depth of the scattering medium supposed to let a fraction of the continuum illuminate the disk  \citep{Chen1989,dumontcollin-souffrin90}.  In this cases, the relevant timescales are the dynamical timescales $t_{\rm dyn} \sim 1/\Omega_{\rm K}$, less than 1 yr at $\sim 1000 r_{\rm g}$, the amount of ionized gas needed to account for the H$\alpha$\ luminosity $M_{\rm HII} \sim 0.13 L_{40.75} n_{\rm H, 9}^{-1}$ M$_\odot$, where$\log L_{40.75}$\ erg s$^{-1}$. The light travel time to $\sim$ 2000 $r_{\rm g}$ is correspondingly short, $\sim 2  $d. Although a more realistic configuration may increase both the light-travel time and the gas mass involved, substantial changes may still occur on relatively short timescales ($\sim 1$ yr), involving only a limited amount of line-emitting gas ($\sim 1$ M$_\odot$).

\begin{figure}
    \centering
    \includegraphics[width=1.0\columnwidth]{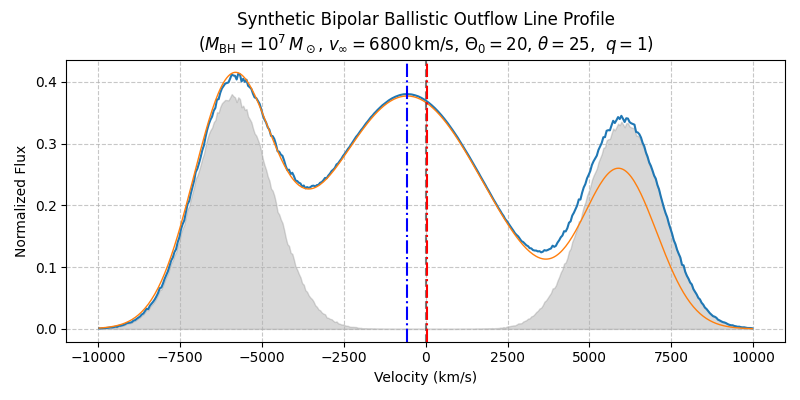}
    \caption{The emission line profile produced by a bipolar outflow constrained between a cone of semiaperture $\Theta_0 = 20$ degrees, with the cone axis inclined by $i = 5$\ with respect to the line of sight (blue line).  The bipolar outflow produces the satellite components that are observed at $\approx \pm 7500$ \kms. A central Gaussian component with the same parameters (FWHM, shift) of the empirical model shown in Fig. \ref{fig:haplotSDSS} is added for easing a comparison. The full empirical model is shown as an orange line.  }
    \label{fig:biconemodel}
\end{figure}

\subsection{A bipolar outflow model}
\label{bicone}

\begin{figure*}
    \centering
    \includegraphics[width=2.08\columnwidth]{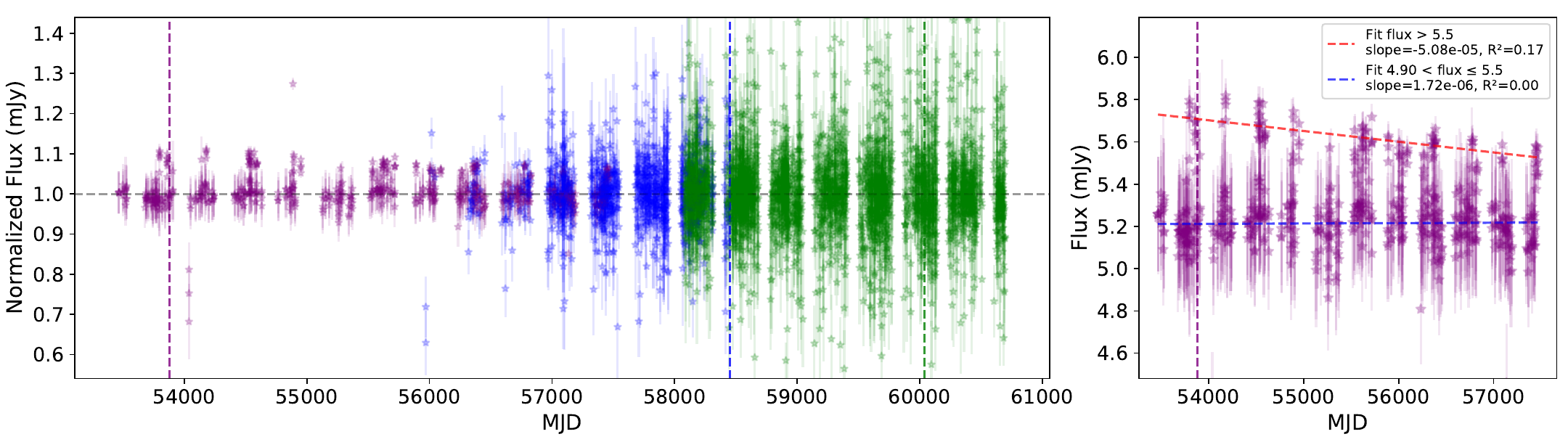}
    \caption{
    Left panel: Optical light curve of UNAM-KIAS~613 from 2005 to 2025 in the $V$ and $g$ bands, with fluxes normalized for comparison. Purple symbols correspond to Catalina data \citep{Drake2014}, while blue and green symbols correspond to ASAS-SN observations \citep{Kochanek2017}. Vertical dashed lines mark the epochs of spectroscopic observations: SDSS (purple), Asiago (blue), and OAN-SPM (green). No significant flaring events are detected over this time span. Right panel: Linear fits to two flux regimes identified in the Catalina $V$-band data. 
    } 
    \label{fig:lightcurve}
\end{figure*}

As discussed in Section \ref{disk}, a  fit to the 2006 H$\alpha$ profile using the relativistic accretion disk model of \citet{Chen1989} yields a satisfactory match, reproducing features such as the asymmetric wings and double-peaked structure, and implying a black hole mass of $\log M \sim 7 $ [M$_{\odot}$]. However, interpreting this result in a broader temporal context suggests caution. Later spectroscopic observations from 2018 and 2023 reveal only a single broad H$\alpha$ component, with no broad H$\beta$ emission in either case. The Eddington ratios during these epochs remain low, but slightly increased ($\lambda_{\rm Edd} \approx 0.040$ in 2018 and $\lambda_{\rm Edd} \approx 0.038$ in 2023 vs. $\approx 0.03$ in 2006). In addition, we recovered photometric data for the $V$-band and $g$-band for UNAM-KIAS 613 that span the period from 2012 to 2025 using the ASAS-SN database. The magnitudes are computed in real time through aperture photometry, with photometric zero points calibrated against the APASS catalog \citep{Kochanek2017}.
The resulting light curve (Figure \ref{fig:lightcurve} left panel) shows no evidence of flaring events or persistent continuum brightening throughout this period in any band, indicating a stable nuclear state. 

Given the isolated nature of this galaxy and the absence of evidence for recent gas inflows, we propose that the broad, multicomponent H$\alpha$ profile observed in 2006 was the result of a one-time  episode. This scenario is consistent with models predicting that, below a critical Eddington threshold, the BLR becomes unstable or disappears altogether \citep{Nicastro2000,Laor2003,Elitzur2009,Trump2011}. In this context, the good fit of the Chen \& Halpern model may reflect a phenomenological match to the transient line kinematics rather than a stable accretion disk structure.

Bipolar outflows \citep{Elvis2000} can produce profiles resembling those of rotating disks, especially in low-Eddington ratio AGNs \citep{Halpern1994,popovicetal14}. A case in point is Pictor A \citep[][]{sulenticetal95} that developed a double-peaked structure not dissimilar from the one of UK 613, from a single-peaked central profile. The previous results point toward a reproducible phenomenon and a common origin for the phenomenology of Pic A and UK 613. A toy model assuming  a biconical geometry and a ballistic velocity field with a cone-half opening angle $\Theta_0 = 20$, angle between the cone axis and the line of sight $\theta = 25 $, $M_\mathrm{BH} = 10^7$ M$_\odot$, launching radius $\approx 3 \cdot 10^{15}$\ cm, terminal velocity  6800 \kms, local velocity dispersion $\sigma \approx 1000$ \kms\ accounts for the width and shift of the bicone features (Fig. \ref{fig:biconemodel}). The model included Doppler boosting and gravitational redshift. The blue-to-red intensity ratio is larger than predicted by Doppler boosting of the approaching side, and de-boosting of the receding one. The red peak weakness could arise by extinction, if those photons originate on the far side of the black hole  where they are farther from the observed, and should pass through more intervening material. 

An AGN with $\lambda_\mathrm{Edd} \sim$-0.01 is possibly in a radiatively inefficient regime where the accretion flow may be in part advection dominated \citep[][i.e., an ion torus]{narayanyi94}. The BLR might be  weak but can still exist, giving rise to very broad emission line profiles. Outflows driven by radiation pressure such as the ones observed in highly accreting quasars \citep[e.g., ][]{martinez-aldamaetal18} are highly unlikely, but mechanical driving (jet-related) becomes important \citep{narayanyi95}. Observing a sudden bipolar outflow producing double-peaked profiles suggests a mechanical outflow or jet–BLR gas interaction rather than classical radiation-driven disk winds \citep{birminghametal25,meyeretal25}. A low-power radio jet can produce localized outspurts or plasma blobs that could propagate along the axis at relativistic or mildly relativistic speeds, encounter clumpy BLR gas, shocking, compressing, or entraining it, and finally accelerate clouds or ionized filaments along the jet direction \citep{normanmiley84}.   Such entrainment can produce bipolar velocity signatures (double peaks) if the jet is oriented close to the line of sight, short-lived  enhancements in line emission due to shock ionization or compression, a transient double-peaked component superimposed on the usual low-luminosity BLR emission. As the shocked gas cools, it remains however exposed to the AGN ionizing continuum, producing significant line emission.

\subsection{A tidal disruption event?}

Tidal Disruption Events (TDEs) occur when a star is torn apart by the gravitational field of a supermassive black hole \citep{komossa15}. In the aftermath, the stellar debris can form an accretion disk, often exhibiting broad emission lines. Some of these lines show double-peaked profiles, a spectral signature commonly associated with rotating disk structures. A case in point is ZTF18aarywbt where the H$\alpha$ profile changes bear a striking similarity with the evolution of UK 613 \citep{wardetal24}. 
  
The double-peaked emission profile is a classic signature of line formation in a relativistic Keplerian accretion disk. For UK~613, the evolution from a double- to a single-peaked H$\alpha$ profile could, under the TDE scenario, be interpreted as stellar debris that initially settled into a disk-like configuration and then faded on a timescale $\ll 12$ yr. The separation of the peaks points to an accretion disk or ring viewed at an intermediate inclination angle ($20^{\circ} \lesssim \theta \lesssim 60^{\circ}$), where both Doppler-shifted sides are visible.  

The fallback time   is the time it takes for the most bound debris from the disrupted star to return to pericenter and start accreting. The canonical expression for the fallback time is \citep{rees88,phinney89}:
$$
t_{\text{fallback}} \approx 41 \left( \frac{M_\mathrm{BH}}{10^6 M_\odot} \right)^{1/2} \left( \frac{R_*}{R_\odot} \right)^{3/2} \left( \frac{M_*}{M_\odot} \right)^{-1} \text{ days}
$$
{where  $M_*$ is the mass of the star,  and $R_*$  is the radius of the star. The  $t_{\text{fallback}}$   can be compared to the orbital time scale} 
  $$
  t_{\text{orb}} = 2\pi \sqrt{\frac{R^3}{G M_\mathrm{BH}}}
  $$
The shortest return time (i.e., of the fallback) should  be $\le$ orbital timescale of innermost emitting material. The point worth stressing here is that and orbital time $t_\mathrm{orb} \approx 100$d derived from the parameters of UK 613, is consistent with a fallback time $\approx 70$d. Both timescales are rather short and point-out toward a short-lived phenomenology of timescale $\lesssim 1$ yr.

An additional clue comes from the Catalina survey light curve (Figure~\ref{fig:lightcurve}, purple stars, right panel), where two groups of data can be distinguished. A linear fit shows one set remaining approximately stable at $5.2 \pm 0.3$ mJy, while the other displays systematically higher values that decline with time. The convergence of this second group toward the stable level, measured since the epoch of the SDSS spectrum (MJD 53879, when the double peak was detected), corresponds to a timescale of 16--22 years, which is significantly longer than the expected duration of a typical tidal disruption event. We therefore caution that these apparent trends may partly arise from instrumental or calibration systematics. In addition, the ASAS-SN light curves are too noisy to reveal comparable features in more recent years, preventing a definitive test of this behavior.
 
In a TDE the stellar debris does not automatically circularize into a clean accretion disk. Whether circularization happens efficiently depends on several physical conditions. Numerical simulations show that circularization can be slow or incomplete. \citep{hayasalietal16}. Even in the case a ring-like configuration is achieved,  double-peaked profiles are short lived \citep[on a timescale of a few months,][]{charalampopoulosetal22}. 

\subsection{Qualitative Bayesian inferences}

A Bayesian interpretation of the available evidence favors models that can naturally reproduce a genuinely  double-peaked broad-line profile in 2006 while accommodating its disappearance by 2018–2023 without a comparably strong optical flare or major secular continuum change. In this framework, the posterior probability of each model given the data, $P(M_\mathrm{i} \mid D)\propto P(D \mid M_\mathrm{i})P(M_\mathrm{i})$, reflects both its prior plausibility $ P(M_\mathrm{i})$ and its likelihood $P(D \mid M_\mathrm{i})$ of the observed profile shape, spectroscopic evolution, and photometric behavior for each model. The accretion-disk  interpretation is  supported by the line profile itself, since classical double-peaked Balmer profiles are a natural outcome of line emission from a rotating disk, although the complete disappearance of such a profile appears to be uncommon among known disk emitters. A point against the disk interpretation is the persistence of the central BLR component, implying that the event might not a global disappearance of the BLR or even the entire disk, but rather the fading of an additional kinematic component superposed on an otherwise normal BLR. A TDE gains support from the transient nature of the phenomenon, but just a few TDEs are now known to exhibit disk-like or even double-peaked Balmer emission \citep{hungetal20}, and they are rather short lived. In the present case, the TDE  likelihood is further reduced by the absence of a strong flare and by the close resemblance of the 2006 spectrum to a classical AGN disk-emitter profile rather than to a short-lived luminous transient. A bipolar outflow triggered by a change in accretion mode remains a viable alternative because it is consistent with a transient profile evolution (expected especially at low Eddington ratio $\sim 0.01$), and its plausibility remains high since the 2006 line can be reproduced by an outflow kinematic signature even more successfully than by a disk model.  However, this possibility is weakened by the lack of strong accompanying continuum dimming or other known, clear state-change tracers. Thus, under reasonable priors, the current evidence appears to favor a transient outflow/state-change scenario with a disk-emitter disappearance  remaining competitive and a TDE explanation less likely.

\section{Multiwavelength Diagnostics of the Ionizing Source in UNAM-KIAS 613}
\label{ref:multiwavelenght}

UNAM-KIAS 613 has been photometrically classified as a blue elliptical  \citep{Lacerna2016}, a designation often associated with residual or recent star formation in early-type systems. However, its spectroscopic and multiwavelength properties strongly suggest the presence of an active nucleus. To constrain the nature of its emission, we compiled luminosities across the optical, mid-infrared (MIR), and radio regimes, and compared its placement in various diagnostic diagrams.

The mid-infrared properties derived from WISE photometry \citep{Wright2010} provide a first indication of AGN activity. Using Vega-calibrated magnitudes in the W1–W4 bands, we computed monochromatic luminosities (Table~\ref{tab:multiwavelenght}) and found that the spectral energy distribution (SED) peaks at 3.4\,$\mu$m (W1), with a secondary maximum at 22\,$\mu$m (W4), and lower values at 4.6\,$\mu$m (W2) and 12\,$\mu$m (W3). This SED shape is atypical of galaxies dominated by thermal dust emission, such as luminous infrared galaxies (LIRGs) or starbursts, where the peak typically occurs at longer wavelengths (10–30\,$\mu$m) due to UV photon reprocessing by warm dust \citep{Jarrett2011}. Instead, the MIR emission of UNAM-KIAS 613 is more consistent with a compact AGN continuum, likely associated with hot dust near the sublimation radius or non-thermal coronal processes. The MIR luminosities, with \( \log(\nu L_\nu\ [\text{erg\,s}^{-1}]) \approx 42.7\text{–}43.1 \), exceed those of quiescent ellipticals but remain below the values typical of powerful AGNs or starbursts, supporting its classification as a low-luminosity AGN \citep{Mateos2013,Assef2013}.

\begin{figure}
    \centering
    \includegraphics[width=\columnwidth]{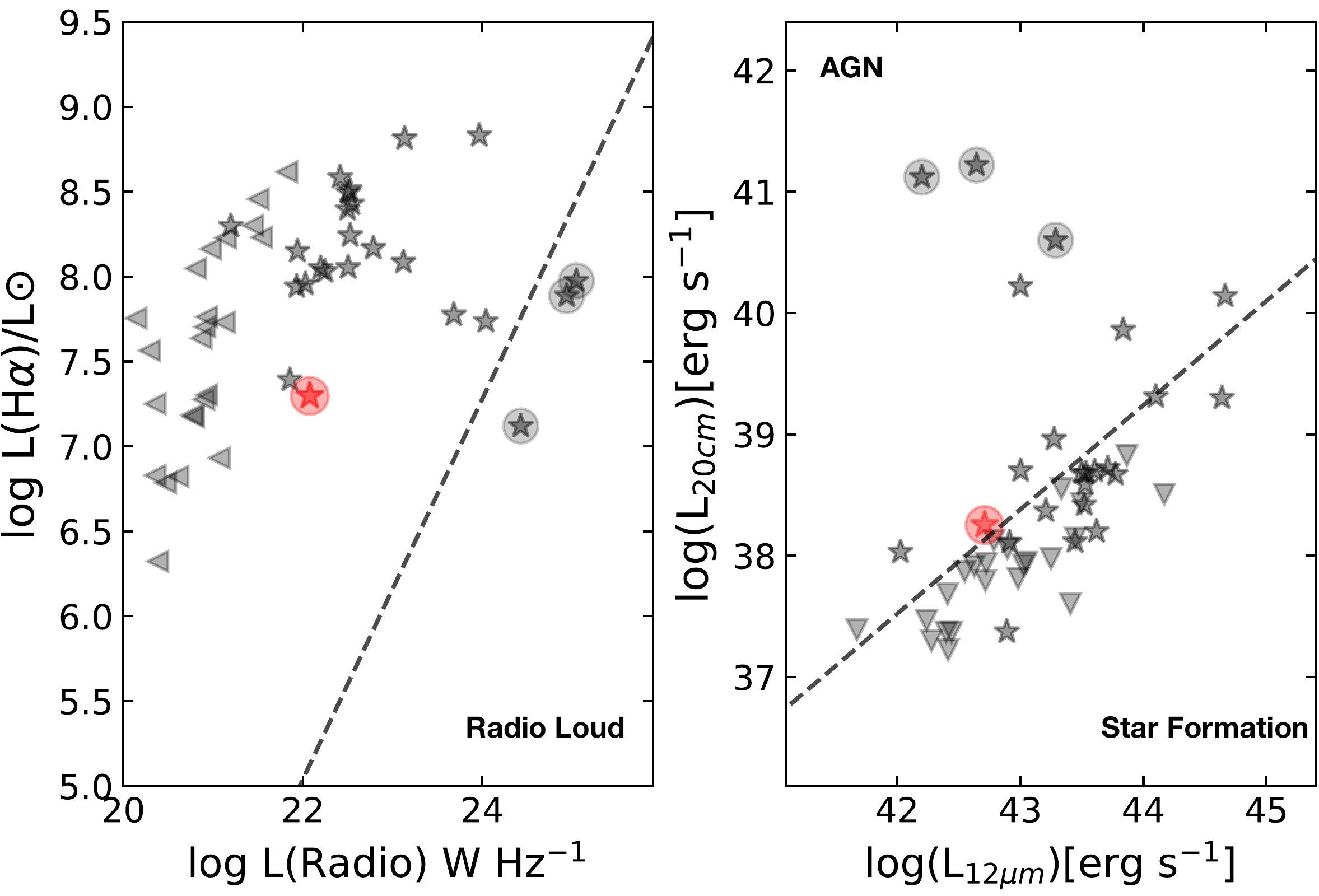}
    \caption{The classification of radio sources. Left plot shows the limit between radio loud and star forming sources \citep{Best2012} while right plot the best fit of star forming sources from \citet{Mingo2016}. UK 613 is shown as red star and the gray markers are from the type 1 AGN sample from \citet{Cortes2022}.}
    \label{fig:Mingo}
\end{figure}

The radio properties provide complementary constraints. From the NVSS survey, UNAM-KIAS 613 is detected as a compact 1.4\,GHz source with a luminosity of \( \log_{10}(L_{1.4\,\mathrm{GHz}}\, [\mathrm{erg\,s^{-1}}]) = 38.17 \), or equivalently \( \log_{10}(L_{1.4\,\mathrm{GHz}}\, [\mathrm{W\,Hz^{-1}}]) = 21.99 \). When compared with the luminosity of the narrow H$\alpha$ component, \( \log_{10}(L_{\mathrm{H}\alpha}/L_\odot) = 7.297 \), the galaxy falls along the locus of star-forming galaxies defined by \citet{Best2012}, suggesting that star formation could contribute significantly to the observed radio emission. However, the diagnostic plane proposed by \citet{Mingo2016}, which compares radio luminosity with MIR W3 (12\,$\mu$m) luminosity, places UNAM-KIAS 613 near the boundary separating AGN- and star formation-dominated systems (see Figure \ref{fig:Mingo}). This indicates that while star formation likely plays a role, a contribution from AGN-related processes cannot be excluded.

\begin{figure}
    \centering
    \includegraphics[width=\columnwidth]{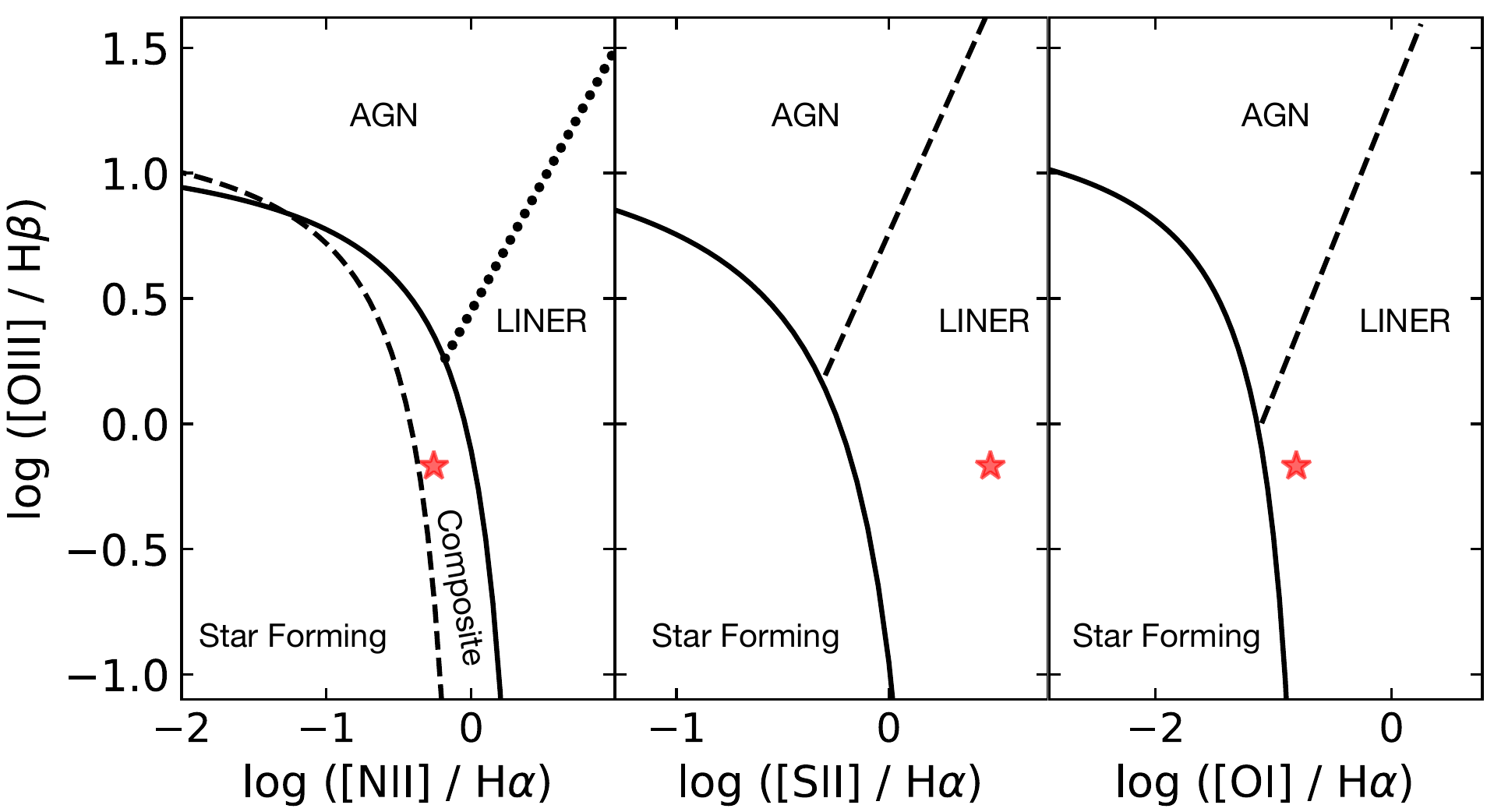}
    \caption{The narrow line diagnostic diagrams from \citet{Baldwin1981}. UK 613 is shown as a red star.}
    \label{fig:BPT}
\end{figure}

Further constraints come from optical emission-line diagnostics. Based on the BPT diagrams \citep[][Figure~\ref{fig:BPT}]{Baldwin1981}, and adopting the demarcation curves of \citet{Kewley2001,Kewley2006,Schawinski2007,Kauffmann2003}, we analyzed the narrow emission-line fluxes measured from the SDSS spectrum after subtracting the host-galaxy contribution. UNAM-KIAS 613 falls in the LINER region of the [O\,\textsc{i}]/H$\alpha$ and [S\,\textsc{ii}]/H$\alpha$ diagrams, while in the [N\,\textsc{ii}]/H$\alpha$ diagram it lies in the composite region, consistent with a mixture of ionization from both star formation and AGN activity. This classification supports the interpretation of UNAM-KIAS 613 as a transition object, in which nuclear activity dominates but residual star formation is still present.

Altogether, the multiwavelength evidence converges toward the conclusion that UNAM-KIAS 613 hosts a low-luminosity AGN. While some contribution from star formation is likely—particularly in the radio and optical narrow-line emission—the MIR SED, the BPT classification, and its placement in radio–MIR diagnostic diagrams favor an AGN origin for the bulk of its ionizing power.

\begin{table}
    \centering
    \begin{tabular}{ccccccc}
    \hline
    \hline
    Spectrum & w1 & w2 & w3 & w4 & 20 cm & $\rm H\alpha_{NC}$ \\
    (1) & (2) & (3) & (4) & (5) & (6) & (7) \\
    \hline
    SDSS & 43.07 & 42.77 & 42.71 & 42.85 & 38.17 & 40.90 \\
    Asiago & - & - & - & - & - & 40.48 \\
    OAN-SPM & - & - & - & - & - & 40.98  \\

    \hline
    \end{tabular}
    \caption{The multiwavelength $\rm\lambda$ L$\rm_\lambda$ luminosities (in log erg s$^{-1}$) of UNAM Kias 613. Columns 2–5 are the \textit{WISE} luminosities for each Near-IR and MID-IR band. Column 6 is the radio luminosity (L$\rm_{Radio}$) at 20 cm from the NVSS catalogue. Column 7 is the $\rm H\alpha$ narrow component luminosity.}.
    \label{tab:multiwavelenght}
\end{table}

\subsection{SEDs comparison}

\begin{figure*}
    \includegraphics[width=2.1\columnwidth]{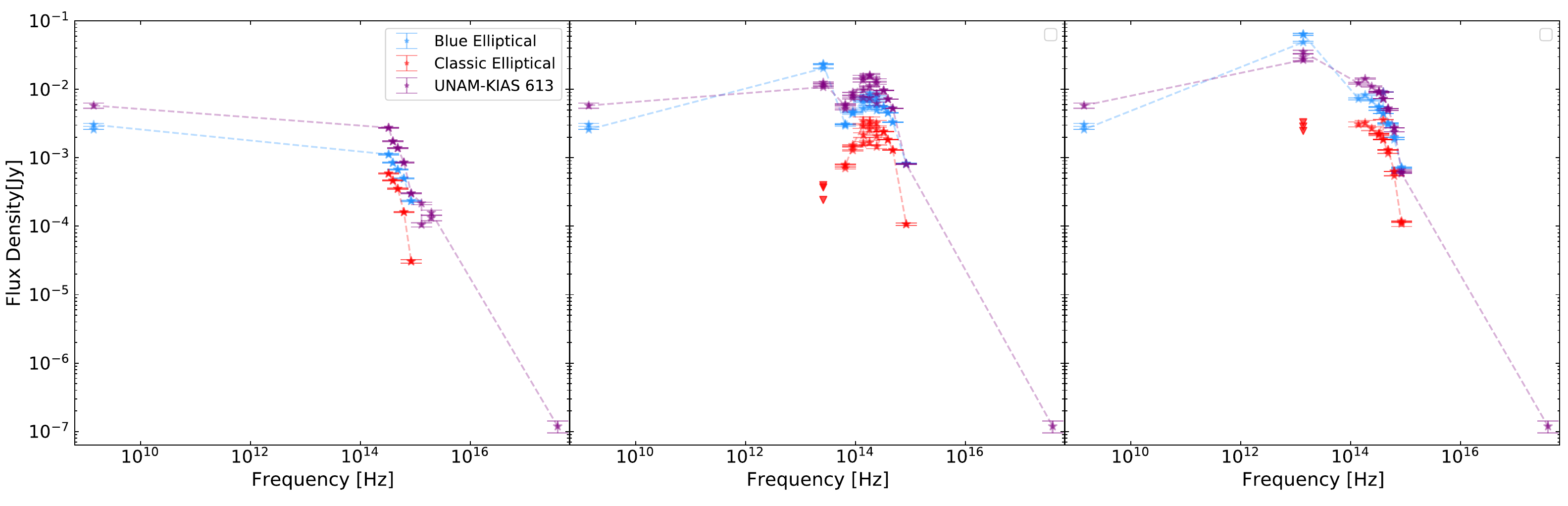}
    \caption{ Spectral Energy Distributions (SED) of three elliptical galaxies. From left to right: SED of small aperture ($\sim$2–5''), SED of medium aperture ($\sim$4–8''), and SED of big aperture (>10''), respectively. In all panels there are the SED of a massive classic elliptical galaxy (red stars), a blue-like less massive elliptical galaxy (blue stars) and UNAM-Kias 613 (purple stars).}
    \label{fig:SEDs}
\end{figure*}

With all the multiwavelength data obtained for UK 613, we constructed its Spectral Energy Distribution (SED), shown in Figure~\ref{fig:SEDs} as the purple line. For comparison, we also include the SED of a massive, classical elliptical galaxy (MaNGA 1-152034; red line) and that of an elliptical galaxy with the same stellar mass as UK 613 but slightly bluer in color (MaNGA 1-153938; blue line). Neither of the comparison ellipticals shows any AGN signatures in their spectra, such as broad or narrow emission lines. 

{ The most relevant SED data to constrain the AGN contribution are in the radio and the X-ray. There is a tremendous degeneracy in the origin of X-ray \citep{netzer13,beckmannshrader12} and weak radio emission \citep{panessaetal19,radcliffeetal21} that cannot be resolved by single-epoch observations. However, some meaningful constraints on the origin of the X-ray emission can be derived by considering the ratio between the X-ray emission and the radio and UV emissions reported in Table \ref{tab:mf_data}. From the specific flux ratio of the X-ray and UV, we derive an a spectral index $\alpha_\mathrm{oX} \approx -1.264$. This value  can be compared with the spectral index derived from the XMM-Cosmos survey. Considering the correlation between $\alpha_\mathrm{oX}$\ and Eddington ratio \citep{lussoetal10}, the  $\alpha_\mathrm{oX} \approx -1.264$ measured for UK 613 yields an Eddington ratio $\approx 0.05$, consistent with the estimation from optical luminosity and bolometric correction. The consensus view is that most of the hard X-ray emission in radio-quiet Type 1 AGN originates in a compact, hot corona of thermal electrons overlaying the accretion disk \citep[e.g.,][]{doneetal12}. The  $\alpha_\mathrm{oX}$\ value of UK 613 is consistent  with an energy output   at energies 2--10 KeV of the optically thin X-ray corona  that typically carries only a few per cent of the AGN bolometric luminosity, whereas the optical–UV disk component contributes on the order of 30–50 per cent of the bolometric emission \citep{kubotadone18}.} 

 {UK 613 is radio-detected, implying a radio power that is relatively modest $\sim 10^{29}$ erg s$^{-1}$ Hz$^{-1}$, and  a radio-to-optical specific flux ratio $R \approx 7$. Such weak radio emission can have multiple origins; however, if we consider the   ratio between the Einstein Observatory X ray flux at 2 KeV and the radio at 1.4 GHz \citep{terashimawilson03}, we obtain $\log R_\mathrm{X} =\nu f(\nu)$(1.4GHz)/$ / \nu f_{\nu} $(2 keV)$\approx$ -3.75. Assuming a spectral index -0.5 in the radio, the same ratio computed at 5 GHz would be  $\log R_\mathrm{X} \approx -4.03$.   This value is typical for radio-quiet  type 1 AGN and QSOs selected from the 2XMMi XMM-Newton X-ray source catalogue \citep[][see their Fig. 8]{balloetal12}. The energy flux ratio $\log R_\mathrm{X}$  of the jet in radio-loud sources is higher, in the range between  1-10 \citep{harriskrawczynski06}.   In summary, aside from the fact that the X-ray, UV, and radio measurements were not taken simultaneously, the available multiwavelength data are   consistent with a low–Eddington‐ratio, radio-quiet Type 1 AGN.  }

\begin{table} 
    \centering
    \begin{tabular}{lccl}
        \hline
        Range  & Frequency (Hz) & Specific flux (Jy) & Reference \\
        \hline
        X-ray    & 3.87E+17       & 1.19E-7   & \citet{gioiaetal90}      \\
        Radio    & 1.40E+9        & 5.80E-3   &  \citet{condonetal98}     \\
        Galex    & 1.29E+15       & 1.61E-4   &  \citet{bianchietal17}     \\
        \hline
    \end{tabular}
    \caption{Frequency and  specific fluxes of key multifrequency observations}
    \label{tab:mf_data}
\end{table}

\section{The Large Scale Environment in the neighborhood of UK 613} \label{LSS-Environment}

In the previous sections, our analysis of the broad emission-line profiles in UK~613 has shown that the observed double-peaked structure is best interpreted as a transient event rather than a stable, long-lived feature. Various scenarios were considered at the nuclear scale, including disk emission, bipolar outflows, and the tidal disruption event hypothesis, all of which point to a short-lived origin for the complex H$\alpha$ profile observed in 2006.  

In this section, we broaden the scope of our study and investigate the large-scale environment of UK~613 to explore whether external factors could have influenced its nuclear activity. Using SDSS survey data, we examine the galaxy’s location within the cosmic web and assess the potential role of its surrounding environment in shaping its spectroscopic properties.  

Previous studies of AGNs in low-density regions have found mixed results regarding the imprint of large-scale structure on nuclear activity. For example, \citet{Constantin2008} reported a higher incidence of AGN in voids than in denser environments, particularly among moderately massive hosts, while \citet{Liu2015} found comparable abundances of AGN in voids and walls. More recently, \citet{Ceccarelli2022} presented statistical evidence for an enhanced fraction of both AGN and star-forming galaxies in voids, regardless of classification scheme (BPT, WHAN, or WISE), suggesting that the void environment can significantly impact AGN fueling and galaxy evolution. Furthermore, interactions between galaxies and their embedding cosmic web structures (e.g., walls or filaments) have been proposed as potential triggers of nuclear activity \citep[][see also \citealt{Benitez13}]{Aragon19}.  

\begin{figure*}
    \centering
    \includegraphics[width=\textwidth]{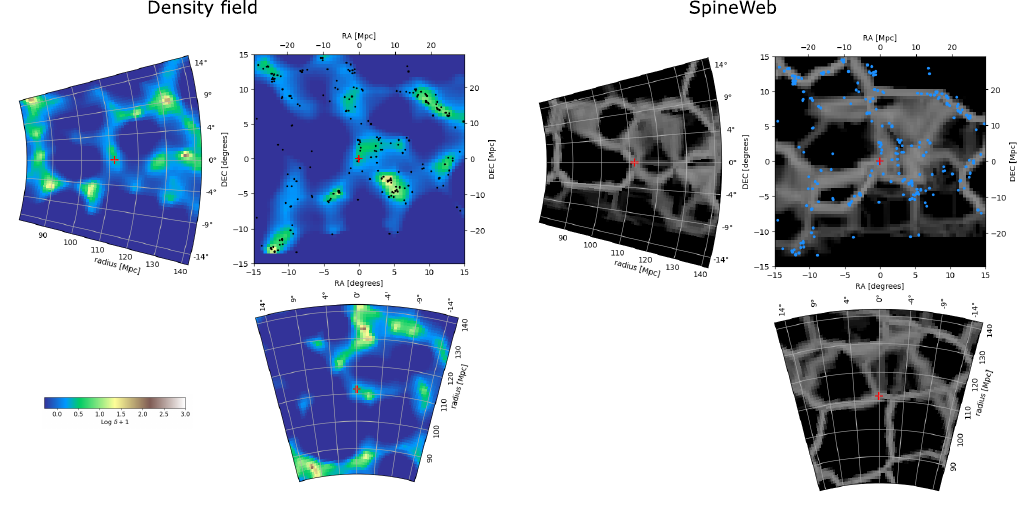}
    \caption{The Large Scale Structure around UK 613. Left: three orthogonal cuts (distance-RA, RA-DEC and RA-distance clockwise) in the density field centered on UK 613 (marked by a red cross) showing the surrounding network of filaments, walls, and voids. The color palette is such that white and intense yellow roughly represents dense regions in clusters and groups, green walls and filaments and dark blue underdense voids. The black dots in the RA-DEC slice correspond to galaxies in a slice 4 Mpc thick along the line of sight and centered at UK 613. Right: same as left panels but here we show the network of voids identified with the SpineWeb method (see text for details). Note the complexity of the environment around UK 613 and the large void right next to it.\label{fig:LSS}}
\end{figure*}

It thus appears suggestive that differences in galaxy growth and properties imprinted across their large-scale environment should also extend to the properties and activity of the central black holes residing within them and their associated AGNs. Following these ideas, in this final section, we utilize the SDSS survey to estimate and analyze the large-scale structure environment around UK 613, following \citep{Aragon07, Aragon2010,Aragon24} with the aim of identifying possible distinctive effects that the large-scale environment may have imprinted, which can help constrain the observed properties.

Fig. \ref{fig:LSS} shows the cosmic structures in the neighbourhood of UK 613. In order to get a better impression of the 3D structures we used a tool developed to show orthogonal cuts in the density field around a given galaxy (Aragón-Calvo in preparation). The density field was computed using the Stochastic Delaunay Tessellation Field Estimator method \citep{Aragon21}, which improves the Delaunay Tessellation Field Estimator algorithm \citep{Schaap00} producing a smooth reconstruction, particularly in low density regions prone to the tessellation artifacts of the original method. The density field around UK 613 is or the order of $delta+1 \sim 2$ which is consistent with either a tenuous filament or a wall. The distance-RA slice shows UK 613 located to the left of a large void ($\sim$ 20 Mpc in diameter) and at the edge of a roughly planar configuration on the plane of the sky (as seen in the RA-DEC slice).

A more explicit view of the surrounding cosmic web is shown in the right panels of Fig. \ref{fig:LSS} where we show the network of walls in three orthogonal slices. The walls were identified using the SpineWeb method \citep{Aragon2010, Aragon24} which segments the density field into voids, walls and filaments following the critical lines in the density field via the watershed transform \citep{Platen07}. The network of walls confirms our initial assessment of the location of UK 613 at the edge of a large planar structure and also at the edge of a large void. 

\begin{figure}
    \centering
    \includegraphics[width=0.8\columnwidth]{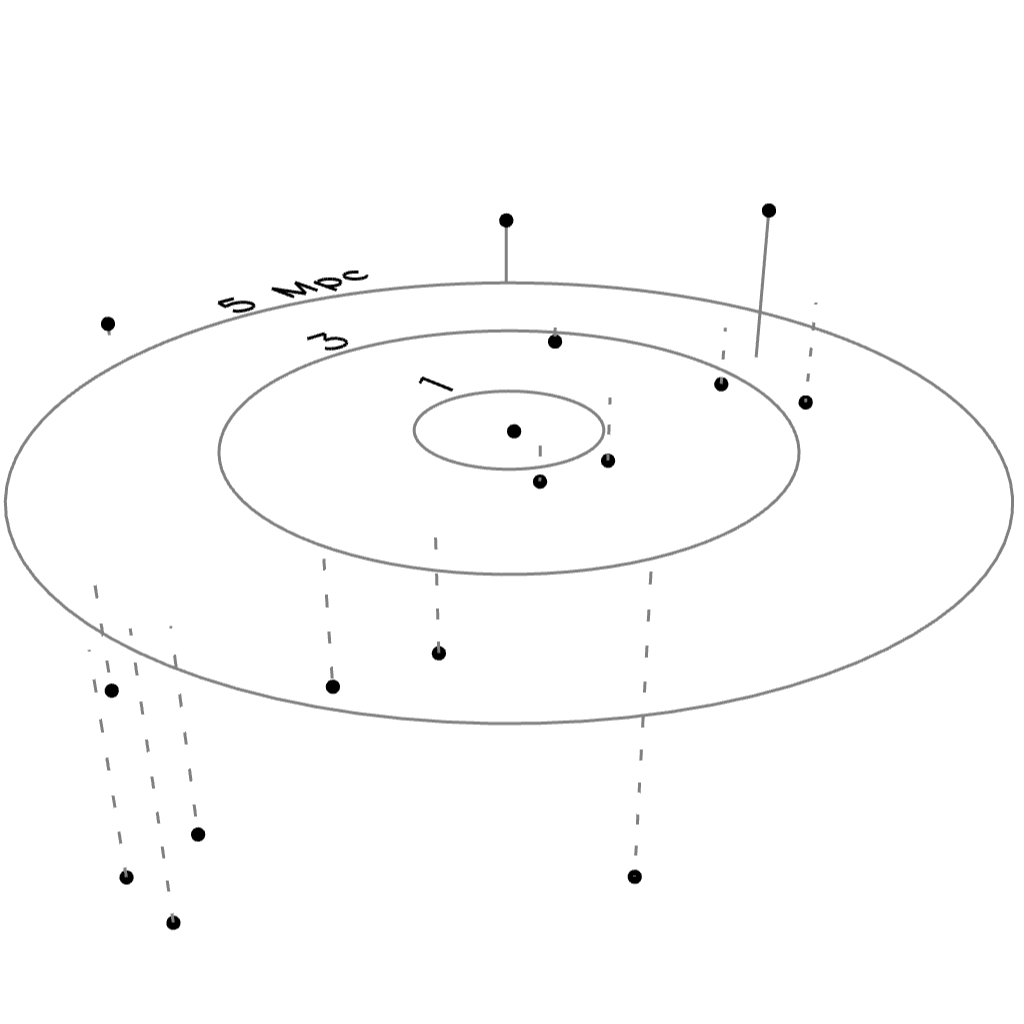}
    \caption{Distribution of galaxies centered on UK 613. Black circles correspond to neighboring galaxies. The horizontal rings show distances of 1,3 and 5 Mpc and the vertical lines indicate the distance from the horizontal plane to each galaxy (solid/dashed lines correspond to galaxies above//below the plane respectively). For visualization purposes the galaxies were oriented with the local geometry (see text for details).\label{fig:galaxies}}
    
\end{figure}

Figure \ref{fig:galaxies} shows the actual galaxy distribution around UK 613 inside 6 Mpc. For visualization purposes the galaxy positions were aligned with the eigenvalues of the inertia tensor in order to highlight the local geometry. The 3D galaxy distribution shows that UK 613 is located inside a thin filament (running from the bottom-left to the top-right in the figure) that is part of a larger planar association (as seen in the RA-DEC slices in Fig. \ref{fig:galaxies}. 

UNAM-KIAS 613 is an example of the complex interplay of an accretion disk, outflows, and host-galaxy processes in low-accretion, low-mass black hole AGN, hosted in a blue early-type galaxy of intermediate stellar mass, that resides in a large-scale wall structure. Although initially classified as a Seyfert 1.8, the actual spectral properties of UK 613 place it as a low-luminosity LINER or transition object, with a modest Eddington ratio (around 0.05) and weak radio emission typical of radio-quiet AGNs.

The observed long-term spectroscopic evolution of UK 613 described in the present work, emphasizes however, the need for additional processes to explain its behavior, including episodic fueling or perturbations caused by close neighbors within its large-scale environment.

Figure \ref{fig:galaxies} emphasizes the absence of similar-sized galaxies within the first Mpc, consistent with the selection criteria applied in the compilation of UK 613 (see Section 2). Notice however, that although this condition avoids the effects of major mergers in the last few gigayears, another intervening condition, namely, the magnitude gap ($\Delta m_r$) $\geq 2.5$ mag, does not exclude the presence of smaller galaxies which could be a possible source of local perturbations in such isolated galaxies like UK 613. This is consistent with other studies looking for the effects of the local environment on the AGNs phenomena \citep{Kauffmann2004,Choi2009,Coldwell2014,Coldwell2017}.

More recently, and in line with our results, \citet{Jaber2024} employed the SpineWeb method to investigate the hierarchical nature of substructures within the cosmic web, examining their impact on the properties of galaxies in voids. They emphasize that cosmic voids possess an intricate internal network of substructures, making them a complex environment for galaxy formation, which impacts the properties and evolution of the galaxies that form within them in a unique way (see also \citealt{Aragon13}). 

An interesting scenario for cosmic-web triggered AGN gas accretion is the interaction between galaxies and their host LSS structure. \citet{Benitez13} presented a study of dwarf galaxies in walls and show that they interact with their host wall losing gas in the process. This could perturb the gas inside the galaxy, triggering gas accretion by its central black hole.

\section{Summary and Conclusions} 
\label{Conclusions}

Unlike Arp~102B, where disk-related broad profiles remain relatively stable, UNAM-KIAS~613 exhibits the complete disappearance of disk-like double-peaked components while a central broad-line component persists within just over a decade. This drastic change points to an unusual evolutionary event in the structure of its broad-line region, not commonly observed in disk-emitting AGN, and highlights UNAM-KIAS~613 as an almost uncommon case of broad-line spectral evolution in a low-luminosity AGN.  

Multiwavelength diagnostics reinforces this interpretation. The mid-infrared spectral energy distribution derived from WISE photometry peaks at short wavelengths, consistent with hot dust heated by an AGN rather than star formation. In the radio regime, UNAM-KIAS~613 lies near the boundary between AGN- and star formation–dominated populations depending on the diagnostic plane, while optical BPT diagrams classify it as a LINER in [O\,\textsc{i}] and [S\,\textsc{ii}], but composite in [N\,\textsc{ii}], indicating mixed ionization from both AGN activity and residual star formation. Together, these diagnostics establish UNAM-KIAS~613 as a transition system, where nuclear activity coexists with low-level star formation.  

To explore the origin of the transient double-peaked profile, we performed multiple fits and explored several models, including Gaussian decompositions, disk-emitter scenarios, and bipolar outflows, as well as testing the tidal disruption event hypothesis. The temporal behavior of the profile, with the double-peaked structure present only in the earliest epoch, and absent more than ten year after, is poorly constrained. It favors a one-time bipolar outflow but leaves open  interpretations such as  a TDE or the disappearance of the double-peaked emission line emitting region. The  possibility of a TDE and especially of an illumination change cannot be definitively excluded.

Regarding the large-scale environment, UNAM-KIAS~613 illustrates how the spectral evolution of low-luminosity AGN can be influenced by their position in the cosmic web. Although locally isolated, UK~613 lies within a wall at the edge of a large void, where subtle perturbations from small companions or interactions with the cosmic web may provide intermittent fueling. This interplay between environment and nuclear processes offers a plausible explanation for its transient broad-line features and current low-accretion LINER state, though this connection is a working hyphothesis and calls for substantially deeper surveys such as the ones that will be made possible by the proposed Wide Field Spectrographic Telescope \citep{mainierietal24}  — to assess its viability.  

Ultimately, low-luminosity AGN such as UNAM-KIAS~613 sit at the nexus of several competing physical regimes: as the accretion rate drops below a few percent of Eddington, the classical thin disk may give way to a hot, radiatively inefficient flow, suppressing the UV–optical continuum while favoring stronger non-thermal emission. In such systems, additional processes---including episodic fueling, variability of the accretion structure, or perturbations induced by a secondary black hole---can further destabilize the disk, driving structural changes that leave a direct imprint on the broad emission-line profiles.

Overall, UNAM-KIAS~613 exemplifies the complex interplay of disk, outflow, and host-galaxy processes in low-accretion, low-mass black hole AGN. Its evolution from a double-peaked emitter to a single-component profile underscores how dramatic structural changes can occur in the absence of major luminosity variations, highlighting the need for long-term spectroscopic monitoring. Ultimately, our results suggest that UNAM-KIAS~613 underwent a transition between radiatively inefficient and efficient accretion modes, offering a rare window into a poorly explored AGN population.

\section*{Acknowledgements}
\addcontentsline{toc}{section}{Acknowledgements}

We thank the anonymous referee for providing constructive comments that helped to improve the manuscript. 
ECS acknowledges the fellowship 825458 from Secretaría de Ciencia, Humanidades, Tecnología e Innovación (SECIHTI)-México through the program "Estancias Posdoctorales por México". 
HMHT acknowledges the CONAHCYT project CF-G-543. HMHT acknowledges support from CONAHCYT project CF-2023-G-1052. HMHT acknowledges support from SECIHTI project CBF-2025-G-1327.    
PM acknowledges financial support from the Spanish MCIU through project PID2022-140871NB-C21 by “ERDF A way of making Europe”, and from the Severo Ochoa grant CEX2021-515001131-S funded by MCIN/AEI/10.13039/501100011033.
MAAC acknowledges support from CONAHCyT Ciencia de Frontera grant CF-2023-1-1971 and UNAM PAPIIT grant IN115224. 
CAN acknowledge the support from projects CONAHCyT CBF2023-2024-1418, PAPIIT IA104325, IN119123, DGAPA PAPIIT IN-113026 and IN-113726.
Based upon observations carried out at the Observatorio Astronómico Nacional on the Sierra San Pedro Mártir (OAN-SPM), Baja California, México. We thank the daytime and night support staff at the OAN-SPM for facilitating and helping obtain our observations.
Based in part on observations collected at Copernico 1.82m telescope  (Asiago Mount Ekar, Italy) INAF - Osservatorio Astronomico di Padova. 
Finally, we wish to acknowledge with gratitude the guidance, friendship and mentorship of Professor Deborah Dultzin, whose influence  has been invaluable to all the authors of this paper.


\section*{Data Availability}

The Sloan Digital Sky Survey (SDSS; \url{https://www.sdss.org}), WISE (\url{https://irsa.ipac.caltech.edu}), Catalina (\url{http://nesssi.cacr.caltech.edu/DataRelease/}), and ASAS-SN (\url{https://asas-sn.osu.edu}) data used in this work are publicly available from their respective archives. 
Additional spectroscopic observations obtained at the Observatorio Astronómico Nacional San Pedro Mártir will be made publicly available in a forthcoming publication.




\bibliographystyle{mnras}
\bibliography{example}




\bsp	
\label{lastpage}
\end{document}